\documentclass[modern]{aastex631}
\usepackage{hyperref}
\usepackage{xspace}
\usepackage{comment}
\usepackage{amsmath,amsthm,amssymb}
\usepackage{bm} 
\usepackage{graphicx}
\usepackage{longtable}
\usepackage{mathrsfs}

\newcommand\superbit{\textsc{SuperBIT}\xspace}

\newcommand{\ee}[1]{\times 10^{#1}}
\newcommand{\SN}[1]{S/N $> {#1}$}
\newcommand{\Neff}{galaxies arcmin$^{-2}$\xspace}
\newcommand{\umag}{$u$\xspace}
\newcommand{\bmag}{$b$\xspace}
\newcommand{\lum}{$lum$\xspace}
\newcommand{\shape}{$shape$\xspace}

\newcommand\utias{\affiliation{University of Toronto Institute for Aerospace Studies (UTIAS), 4925 Dufferin Street, Toronto, ON, Canada}}
\newcommand\dunlap{\affiliation{Dunlap Institute for Astronomy and Astrophysics, University of Toronto, 50 St. George Street, Toronto, ON, Canada}}
\newcommand\princeton{\affiliation{Department of Physics, Princeton University, Jadwin Hall, Princeton, NJ, USA}}
\newcommand\durhamcfai{\affiliation{Centre for Advanced Instrumentation (CfAI), Durham University, South Road, Durham DH1 3LE, UK}}

\newcommand\uoftastro{\affiliation{Department of Astronomy, University of Toronto, 50 St. George Street, Toronto, ON, Canada}}
\newcommand\uoftphysics{\affiliation{Department of Physics, University of Toronto, 60 St. George Street, Toronto, ON, Canada}}
\newcommand\jpl{\affiliation{Jet Propulsion Laboratory (JPL), California Institute of Technology, 4800 Oak Grove Drive, Pasadena, CA, USA}}
\newcommand\durhamcfea{\affiliation{Centre for Extragalactic Astronomy, Department of Physics, Durham University, Durham DH1 3LE, UK}}
\newcommand\durhamifcc{\affiliation{Institute for Computational Cosmology, Durham University, South Road, Durham DH1 3LE, UK}}

\newcommand\washu{\affiliation{Department of Physics, Washington University in St. Louis, 1 Brookings Drive, St. Louis, MO, USA, 63130}}
\newcommand\mcdonnell{\affiliation{McDonnell Center for the Space Sciences, Washington University in St. Louis, 1 Brookings Dr., St. Louis, MO USA 63130}}
\newcommand\sydney{\affiliation{Sydney Consortium for Particle Physics and Cosmology, School of Physics, The University of Sydney, NSW 2006, Australia}}

\shorttitle{Lensing in the Blue II: Methods and Sensitivity}
\shortauthors{McCleary et al.}

\begin{document}

\title{
Lensing in the Blue II: Estimating the Sensitivity of Stratospheric Balloons to Weak Gravitational Lensing}

\author[0000-0002-9883-7460]{Jacqueline E. McCleary}
\affiliation{Department of Physics, Northeastern University, 360 Huntington Ave, Boston, MA}

\correspondingauthor{Jacqueline E. McCleary}
\email{j.mccleary@northeastern.edu}

\author[0000-0002-3745-2882]{Spencer W. Everett}
\jpl

\author[0000-0002-7600-3190]{Mohamed M. Shaaban}
\uoftphysics{}
\dunlap

\author[0000-0002-3937-4662]{Ajay S. Gill}
\uoftastro{}
\dunlap

\author[0009-0006-2684-2961]{Georgios N. Vassilakis}
\affiliation{Department of Physics, Northeastern University, 360 Huntington Ave, Boston, MA}

\author[0000-0002-9378-3424]{Eric M. Huff}
\jpl

\author[0000-0002-6085-3780]{Richard J.\ Massey}
\durhamifcc
\durhamcfea
\durhamcfai

\author[0000-0002-4214-9298]{Steven J.\ Benton}
\princeton{}

\author{Anthony M.\ Brown}
\durhamcfai
\durhamcfea

\author{Paul Clark}
\durhamcfai

\author{Bradley Holder}
\utias
\dunlap

\author{Aurelien A.\ Fraisse}
\princeton

\author{Mathilde Jauzac}
\durhamifcc
\durhamcfea

\author{William C.\ Jones}
\princeton

\author{David Lagattuta}
\durhamcfea

\author{Jason S.-Y.\ Leung}
\uoftastro
\dunlap

\author{Lun Li}
\princeton

\author{Thuy Vy T.\ Luu}
\princeton

\author{Johanna M.\ Nagy}
\washu
\mcdonnell

\author{C.\ Barth Netterfield}
\uoftastro
\dunlap
\uoftphysics

\author[0000-0001-5101-7302]{Emaad Paracha}
\uoftphysics

\author[0000-0002-9618-4371]{Susan F.\ Redmond}
\princeton

\author{Jason D.\ Rhodes}
\jpl

\author{J\"urgen Schmoll}
\durhamcfai

\author[0000-0002-7542-0355]{Ellen Sirks}
\sydney
\durhamifcc

\author{Sut Ieng Tam}
\affiliation{Academia Sinica Institute of Astronomy and Astrophysics (ASIAA), No. 1, Sec. 4, Roosevelt Road, Taipei 10617, Taiwan}

\shortauthors{McCleary et al.}

\begin{abstract}
The Superpressure Balloon-borne Imaging Telescope (\superbit) is a diffraction-limited, wide-field, 0.5 m,
near-infrared to near-ultraviolet observatory designed to exploit the stratosphere's space-like conditions. \superbit's 2023 science flight will deliver deep, blue imaging of galaxy clusters for gravitational lensing analysis. In preparation, we have developed a weak lensing measurement pipeline with modern algorithms for PSF characterization, shape measurement, and shear calibration. We validate our pipeline and forecast \superbit survey properties with simulated galaxy cluster observations in \superbit's near-UV and blue bandpasses. We predict imaging depth, galaxy number (source) density, and redshift distribution for observations in \superbit's three bluest filters; the effect of lensing sample selections is also considered. We find that in three hours of on-sky integration, \superbit can attain a depth of $b = 26$ mag and a total source density exceeding 40 galaxies per square arcminute. Even with the application of lensing-analysis catalog selections, we find $b$-band source densities between 25 and 30 galaxies per square arcminute with a median redshift of $z=1.1$. Our analysis confirms \superbit's capability for weak gravitational lensing measurements in the blue.

\end{abstract}



\section{Introduction} \label{sec:intro}         


The abundance of galaxy clusters as a function of redshift depends sensitively upon both the geometry of the universe \citep{2001ApJ...560L.111H} and the ongoing mechanism of structure formation via gravitational collapse \citep{2001ApJ...553..545H}. Cluster number counts provide a statistically significant constraint on cosmological parameters, and as the largest particle colliders in the Universe, galaxy clusters themselves are proving grounds for alternative models of dark matter \citep{clowe2004weak}.
    
Because most of the mass in a cluster is invisible dark matter, a major challenge confronting cluster cosmology is the difficulty of measuring their masses. The most direct method takes advantage of clusters' weak gravitational lensing signal: the small but coherent magnification of background galaxy fluxes and observed distortion of background galaxy shapes. High-quality weak gravitational lensing studies illuminate the relationship between the true masses of galaxy clusters and their observable gas and stars.

In this context, our collaboration will deploy the Superpressure Balloon-borne Imaging Telescope (\superbit): a stratospheric imaging system that will deliver space-quality imaging from the near-ultraviolet to the near-infrared. \superbit has been optimized for measurement of cluster gravitational lensing: the telescope has a $15\arcmin \times 23\arcmin$ field of view to enable efficient measurements of the weak lensing signal of galaxy clusters at $z \ge 0.05$, and provides stable, near-diffraction-limited imaging for well-measured galaxy shapes. 

Floating above more than $97\%$ of the Earth's atmosphere, the telescope experiences nearly perfect transmission from 280~nm to 900~nm. The stratosphere also offers low sky backgrounds: \citet{gill2020optical} show that \superbit experiences 23.6--25.5 mag arcsec$^{-2}$ in its $b$ filter (365 nm -- 575 nm), up to three mag arcsec$^{-2}$ fainter than the darkest ground-based sites with 22.7 mag arcsec$^{-2}$.

While most surveys measure weak gravitational lensing at red wavelengths, the dark sky background and diffraction-limited optics in the stratosphere uniquely mean that lensing measurements are more efficient in the blue \citep{2022SPIE12191E..14G,2022AJ....164..245S}.

Beyond weak gravitational lensing measurements, \superbit's deep, blue imaging enables a range of scientific investigations. For example, its near-UV (300-400\,nm) photometry spans the Balmer and 4000\,\AA~breaks used to fit galaxy templates for photometric redshift estimation; including NUV photometry can halve uncertainties on the resulting photometric redshifts \citep{sawicki2019cfht}. 

To prepare for \superbit's 2023 science flight, we have created a suite of simulated \superbit galaxy cluster observations with realistic galaxy flux, size, and redshift distributions. PSF models are informed by previous test flights, and background galaxies are gravitationally lensed by foreground cluster halos. We have then developed a weak lensing analysis pipeline built from modern, publicly-available tools like PIFF for PSF characterization, NGMix for galaxy shape measurement, and Metacalibration for galaxy shear calibration. 

At a basic level, processing the simulated observations validates our pipeline performance. More interestingly, this procedure enables us to flow down science requirements into an efficient observing strategy. 
Galaxy clusters have highly localized weak lensing signal, which makes galaxy number density and average redshift the primary figures of merit for cluster surveys. However, the \textit{total} number density of galaxies observed is less important than the number that survive cuts on redshift, signal-to-noise, and size for weak lensing analysis.
In this paper, we will forecast imaging depths, source density, and redshift distributions for stratospheric observations in \superbit's near-UV and blue bandpasses.

This paper is organized as follows. We summarize the \superbit platform in Section \ref{sec:superbit}, and lensing theory in Section~\ref{sec:wl}.
We describe our galaxy shape measurement pipeline in Section~\ref{sec:pipeline}, and our mock \superbit observations in Section~\ref{sec:the_simulations}. We present our results in Section~\ref{sec:results}, provide additional context in Section~\ref{sec:discussion}, and conclude with Section~\ref{sec:conclusions}.

\section{The \superbit observing platform}\label{sec:superbit} 

\subsection{Instrument}\label{sec:superbit:instrument}
\superbit is a 0.5-m mirror telescope that exploits the super-pressure balloon capabilities provided by the National Aeronautics and Space Administration (NASA), which offers mid-latitude long-duration balloon flights 
up to 100 days.
\superbit has been developed and iteratively improved through four one-night commissioning flights. Successful recovery after each flight enabled efficient, closed-loop engineering cycles. A complete description of the resulting mechanical, thermal, control systems, and software architecture appears in \citet{romualdez2018overview}, \citet{Redmond18}, and \citet{StevenThesis}.

\begin{figure}[htp]
    \centering
    \includegraphics[width=0.8\textwidth]{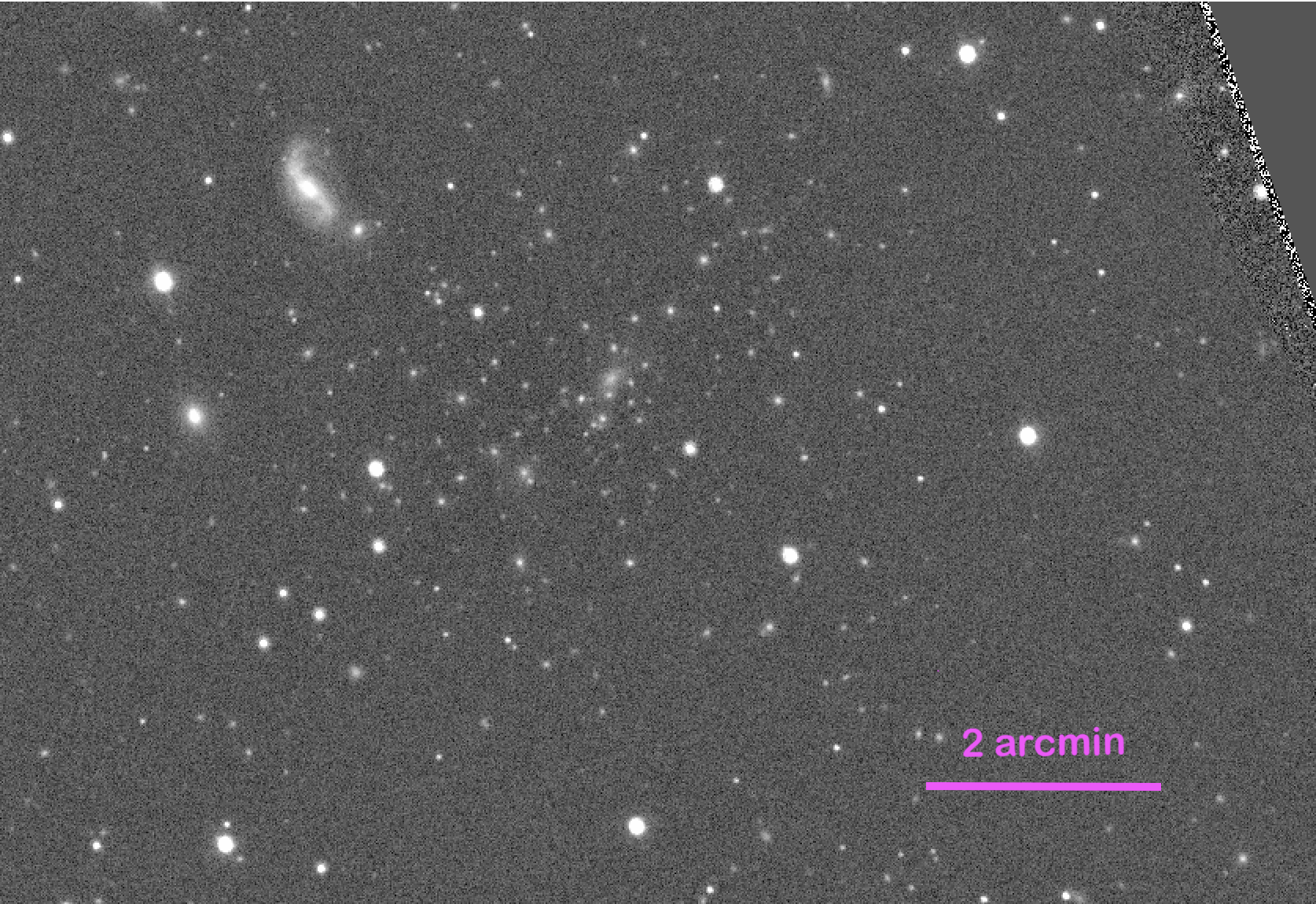}
    \caption{Coadded fifteen-minute lum observation of Abell 2218 made by \superbit during the 2019 engineering test flight.}
    \label{fig:a2218}
\end{figure}

The platform consists of a gondola pointing system and an optical assembly that work together to achieve 0.05\arcsec~focal plane stability via three successive pointing and stabilization regimes: coarse target acquisition to within 0.5\arcmin, fine telescope stabilization at the 0.5\arcsec~level, and finally 0.05\arcsec~image stabilization at the focal plane. 
During the most recent test flight in September 2019, \superbit maintained telescope stability of 0.3\arcsec\ (0.5\arcsec) over a 5~minute (30~minute) exposure, and \textit{image} stability of 0.046\arcsec\ (0.048\arcsec) over a 5~minute (30~minute) exposure. This enabled the first measurements of gravitational lensing from the stratosphere \citep{2020PhDT........17T}, using images of Abell 2218 (Figure~\ref{fig:a2218}), and defined a fiducial exposure time of 5~minutes for future observations \citep{Romualdez_2020}.

\begin{deluxetable}{cccc}\label{tab:skylevels}

\tablecaption{\superbit bandpasses and sky backgrounds}

\tablehead{\colhead{Filter name} & \colhead{Wavelength range} & \colhead{Pivot wavelength} & \colhead{Sky brightness} \\ 
\colhead{} & \colhead{(nm)} & \colhead{(nm)} & \colhead{(e$^-$ s$^{-1}$ pix$^{-1}$)} } 

\startdata
$u$ &  300--435  &  395 &  0.029  \\
$b$ &  365--575  &  476 &  0.052  \\
$g$ &  515--705  &  597 &  0.052  \\
$r$ &  570--720 &  640 &  0.030  \\
$nir$ &  706--1100 & 814 & 0.064  \\
\lum &  370--710  &  522 &  0.084  \\
\shape &  530--830  &  650 &  0.15  \\
\enddata

\tablecomments{Summary of the 2023 flight filters and expected sky brightnesses in each. The $shape$ filter is deprecated and included in this analysis for comparison purposes.
}

\end{deluxetable}

\begin{figure}[htp]
    \centering
    \includegraphics[width=0.8\textwidth]{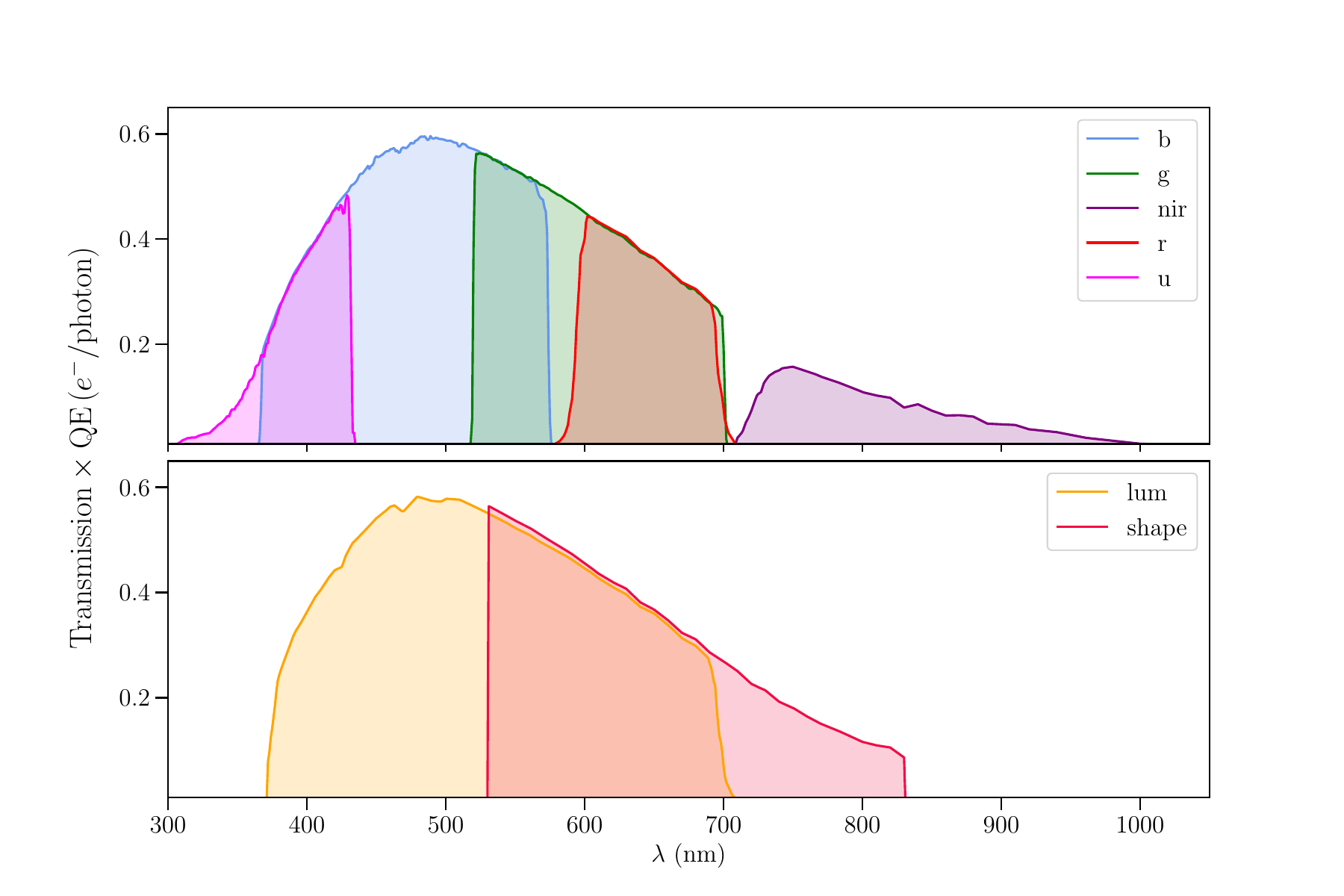}
    \caption{\superbit transmission across the six filters planned for the 2023 flight and the now-deprecated shape band, which will not be flown but is included here for comparative purposes.}
    \label{fig:2023filters}
\end{figure}

Because of the fast development time scales of balloon-borne missions, \superbit has had the ability to upgrade its core technologies between flights. To wit, the 
2023 flight camera is a marked improvement over the CCD flown in 2019. The 2023 science camera is a $9600\times6422$~pixel Sony IMX 455 CMOS detector with 3.76~$\mu m$ (0.141$\arcsec$) square pixels. 
At operating temperature $-10\,\degr$C, this has low read noise (rms $\sim1.7$ e$^{-}$/pixel) and low dark current ($\sim0.0022$~e$^{-}$/s/pixel).
It is sensitive from 300 to 900~nm; its quantum efficiency (QE) and optical throughput are presented in \citet{2022SPIE12191E..14G}. 
Its filter wheel currently includes five broadband filters ($u$, $b$, $g$, $r$, $nir$) 
plus one very broad filter (\lum) designed to collect as much light as possible (Table~\ref{tab:skylevels} and Figure~\ref{fig:2023filters}).
We also show the \shape filter, which is very similar in range to the Euclid VIS filter and was at one point designated for galaxy shape measurements (hence the name).

\subsection{Survey and expected data}\label{sec:superbit:survey_info}

During its planned, up to 100~day science flight from NASA's Long Duration Balloon facility in Wanaka, New Zealand (scheduled for April 2023 at the time of writing), \superbit will be able to observe almost anywhere in the Southern hemisphere and up to 20$^\circ$ North. Target selection will depend on launch date and balloon path, but targets will be automatically drawn from a list of galaxy clusters at redshift $z<0.5$. These include well-studied clusters from the Hubble Frontier Fields, CLASH, RELICS, LoCuSS and COSMOS surveys that are required for calibration, plus merging clusters identified principally via bimodality in \textit{Chandra} X-ray imaging. The clusters have abundant ancillary data: all with X-ray imaging, most with infrared (IR) and radio imaging, and many with substantial investments of ground-based spectroscopy. To these data, \superbit will add deep, wide-field near-UV and optical imaging with angular resolution of $0\farcs3$. With a minimal sample of 45 clusters and assuming a per-cluster scatter in mass of 20\% \citep{2011ApJ...740...25B}, this data can yield weak lensing masses with an ensemble $M_{200c}$ fractional uncertainty of $ 0.2/\sqrt{45} = 0.03$. 
Based on \superbit's 2018 and 2019 test flights, calculations in \citet{2022AJ....164..245S}, and results in this work, each cluster target will be observed for 3 hours (36$\times$300~second exposures) in $b$, plus shorter integrations in $u$ and $g$ bands.


\section{Gravitational Lensing and Metacalibration}\label{sec:wl}


\subsection{Weak gravitational lensing formalism}\label{sec:wl:formalism}

Gravitational lenses like galaxy clusters introduce an isotropic magnification of background galaxies and percent-level distortions in their shapes. The magnification of galaxy images is described by the convergence $\kappa$, a scalar quantity equal to the Laplacian of the gravitational potential of the lens projected along the line of sight. The convergence $\kappa$ can be related to the surface mass density of the galaxy cluster, $\Sigma$, as
\begin{equation}
\kappa \equiv \frac{1}{2} \nabla^2 \Psi(\theta)=\frac{\Sigma}{\Sigma_{\rm crit}};\quad \Sigma_{\rm crit} =\frac{c^2}{4\pi G}\frac{D_{\rm s}}{D_{\rm l} D_{\rm ls}}\label{eqn:sigmacrit}.
\end{equation}
where the critical surface mass density $\Sigma_{\rm crit}$ of the lens depends on the angular diameter distances to the background galaxy $D_{\rm s}$, the lens $D_{\rm l}$ and the lens and source $D_{\rm ls}$, respectively. 

The distortion of galaxy images introduced by gravitational lenses is represented as a complex shear $\gamma$:
\begin{align}
\gamma &= \gamma_1 + i\gamma_2 = |\gamma|e^{2i\phi}
\end{align}

Distortion along the real axes ($x/y$) is described by the $\gamma_1$ component of shear; the $\gamma_2$ component describes the galaxy image distortion along axes rotated through $\pi/4$ radians. The shear $\gamma$ can be related to the cluster gravitational potential $\Psi$ as:
\begin{align}
\gamma(\bm{\theta}) &= \bm{D}\Psi,\\
\bm{D} &= \frac{\partial^2_1 - \partial^2_2}{2} + i\partial_1 \partial_2
\end{align} 

Observations of gravitationally lensed galaxies actually return the reduced shear $\bm{g}$
\begin{equation}\label{eqn:g}
\bm{g} =\frac{\bm{\gamma}}{1-\kappa} = g_1 + i g_2,
\end{equation}
where the variables $g_1$ and $g_2$ in Equation~\ref{eqn:g} are the polarization states of background galaxies with reduced shear $\bm{g}$.

Irrespective of the presence of a gravitational lens, the shapes of galaxies measured on an image can be characterized by an ellipticity $e$:

\begin{equation}\label{eqn:e_vector}
\bm{e} = e_1 + i e_2,
\end{equation}
\begin{equation}\label{eqn:e1_e2}
e_1 = e\,\cos(2\theta), \hspace{0.2cm} e_2 = e\,\sin(2\theta),
\end{equation}
\begin{equation}\label{eqn:e}
e = \frac{a^2-b^2}{a^2+b^2}.
\end{equation}

%
where $a$ and $b$ are the major and minor axes of the galaxy image ellipse. The shear $\gamma$ can be extracted from galaxy ellipticities $e$ in the weak lensing regime, where the distortion introduced by the lens is much smaller than the galaxy images themselves, i.e., where $\kappa,\gamma \ll 1$. In that case, in the absence of intrinsic alignments and for source galaxies at the same redshift, 
\begin{equation}
\gamma \simeq g \simeq \frac{\langle e \rangle}{2 \mathscr{R}}\label{eqn:gamma}
\end{equation}

The factor $\mathscr{R}$ encodes the shear response factor $1 - \sigma^2_{e}$. Because the lensing potential induces curl-free distortions in galaxy images, we estimate the reduced shear about a point on the sky with the tangential ellipticity:
\begin{equation}
g_{\tan} = -(g_1 \cos(2\phi) + g_2 \sin(2\phi)) \label{eqn:gtan}.
\end{equation}

where $\phi$ is the azimuthal angle from the fiducial center of mass to the galaxy. 

Because it is a curl-free statistic, in analogy with electromagnetism, Equation~\ref{eqn:gtan} is sometimes called E-mode signal. A divergence-free statistic, the B-mode or cross-shear, is obtained by rotating Equation~\ref{eqn:gtan} through $\pi/4$ radians:
\begin{equation}
g_{\times} =g_2\cos(2\phi) - g_1\sin(2\phi)\label{eqn:gc}.\footnote{Note that some authors also use the opposite sign convention, $g_{\times} = g_1\sin(2\phi)- g_2\cos(2\phi)$.}
\end{equation}

Galaxy shapes are also convolved with the point spread function (PSF) of the telescope and atmosphere. PSFs tend to circularize galaxy shapes, diluting the real weak lensing signal, while the anisotropic components introduce ellipticities into the galaxy shapes that mimic weak lensing shear. 
Accurate shear inference thus requires that the PSF be modeled and deconvolved from galaxy shape measurements. Readers interested in a comprehensive review of galaxy cluster weak gravitational lensing, including considerations of the PSF, may consult \citet{2020A&ARv..28....7U}.

\subsection{Metacalibration}\label{sec:wl:metacalibration}

In real measurements, the measured galaxy shears $g_1,\, g_2$ are biased estimators of the underlying shear distribution and need to be converted into a true estimator for the weak lensing shear $g_{\rm tan}$. This is generally accomplished by dividing each galaxy's ellipticity by an appropriate ``shear responsivity factor'' $\bm{R}$, which characterizes the response of the galaxy shape estimator $\bm{\hat{g}}$ to an applied shear $\gamma$ \citep{2005PhDT........12H}:

\begin{eqnarray}
    \centering
    \langle \bm{\hat{g}} \rangle &=& \langle \bm{\hat{g}} \rangle |_{\gamma = 0}\  + \langle \bm{R}\gamma \rangle  + O(\gamma^2)\\
    &\approx& \langle \bm{R}\gamma\rangle,
\end{eqnarray}
where in the absence of an external shear field, the average ellipticity should be zero. 

Image simulations are often used to obtain the shear calibration (e.g.~\citealp{mandelbaum2018weak,2017MNRAS.467.1627F}), but face the usual difficulties in replicating all the effects that affect real images. 
Instead, we use the Metacalibration algorithm, which calibrates shear estimators from the galaxy image data itself, without requiring significant prior information about galaxy properties. Metacalibration's data-driven approach is particularly valuable in a new survey like \superbit.

Metacalibration introduces an artificial shear to images and calculates how the shear estimator responds to that applied shear. More specifically, the original galaxy image is deconvolved from the PSF and then sheared by some amount $\gamma$ along each ellipticity component $g_i$. The sheared image is reconvolved with a function slightly larger than the original PSF to suppress noise amplified by the deconvolution process, and measurement of $\bm{\hat{g}}$ is repeated. The shear responsivity $\bm{R}$ is then obtained through the finite difference derivative:

\begin{equation}
\centering
    \bm{R}_{k,l} = \frac{\hat{g}_k^+ - \hat{g}_k^-}{\Delta \gamma_l}
\end{equation}

where $\hat{g}^+$ is the measurement made on an image sheared by $+\gamma_l$ and $\hat{g}^-$ is the measurement made on an image sheared by $-\gamma_l$ for all objects $k$. The responsivities can be computed for every galaxy in an observation catalog, but are very noisy since the ellipticity estimators $\bm{\hat{g}}$ themselves are noisy. 
So in practice, a shear estimate is obtained by dividing the galaxy ellipticity estimator by the mean responsivity over the entire galaxy sample:

\begin{equation}
    \centering
    \langle \bm{\hat{\gamma}}\rangle = \langle \bm{R} \rangle^{-1} \langle \bm{\hat{g}}\rangle
\end{equation}

Estimation of weak lensing shear commonly requires selection cuts on quantities like galaxy size and signal-to-noise ratio.
The probability that a galaxy passes selection cuts changes after the application of an artificial shear. The responsivity then includes both the shear response and the effect of sample selections. We continue to follow the formalism of \cite{sheldon2017practical} and break up the responsivity into two components: 
\begin{equation}
\langle \bm{R} \rangle = \langle \bm{R_\gamma}\rangle + \langle \bm{R}_S\rangle,\label{eqn:resp}
\end{equation}

where brackets denote the average over galaxies $k=1...N_{gals}$, $\langle \bm{R}_\gamma \rangle $ captures the ensemble response of galaxy shapes to an applied shear, and $\langle \bm{R}_S\rangle$ represents the response of the \textit{selections} to an applied shear.


\section{Shape Measurement Pipeline}\label{sec:pipeline}
    

In anticipation of \superbit's 2023 science flight, we have developed a galaxy shape measurement and weak lensing analysis pipeline that employs state-of-the-art algorithms, such as NGMix for optimal estimation of galaxy shapes~\citep{sheldon2015ngmix} and Metacalibration to correct for multiplicative shear bias~\citep{huff2017metacalibration,sheldon2017practical} (also see \ref{sec:wl:metacalibration}). We provide an overview of our pipeline below. Upon acceptance of the paper, we intend to make the pipeline public.

The pipeline is divided into three modules: creation of the input files for galaxy shape fitting (\texttt{medsmaker}); galaxy shape fitting and shear bias correction (\texttt{metacal}); and calculation of the galaxy clusters' tangential and cross shear profiles (\texttt{shear\_profiles}). For ease of use, the pipeline has code to auto-generate the configuration files needed to run this pipeline from beginning to end, based only on a few user inputs. 

\subsection{Creation of shape measurement input files}\label{sec:pipeline:medsmaker}

The input for NGMix and Metacalibration is a multi-epoch data structure (MEDS\footnote{\url{https://github.com/esheldon/meds/wiki/MEDS-Format}}): a kind of FITS binary table with an entry for every object detected in an observation. Each object's MEDS entry contains the following: a postage-stamp cutout of the object, a rendering of the point spread function (PSF) at the location of the object, a weight, a segmentation map, and a bad pixel mask for every exposure in which the object was detected~\citep{Jarvis_2016}.

In our pipeline, MEDS files for \superbit observations are created with the \texttt{medsmaker} module. Much of the ``standard operating procedure'' for astronomical imaging is implemented in \texttt{medsmaker}; we detail the particulars here for reference in future analyses. 

\subsubsection{Detection catalog}

Image data supplied to \texttt{medsmaker} is assumed to be calibrated CMOS or CCD imaging data for which bias subtraction and flat-fielding have already been performed. Exposure weight maps and bad pixel masks are required as well.  

The AstrOmatic tool SWarp is used to combine single-epoch exposures into a deep detection image from which the master observation catalog---the basis of the MEDS file---is obtained with SExtractor \cite{2002ASPC..281..228B}. 
To maximize the number of sources detected, we set a relatively low detection threshold of $1.5\ \sigma$. This necessarily generates spurious detections that do not correspond to galaxies in any exposure. Rather than cut these items out of the MEDS file, which risks introducing an uncontrolled shear selection bias, spurious sources with no cutouts are flagged to be skipped during shape fitting. 
Segmentation maps and catalogs for single-epoch exposures are also generated with SExtractor; segmentation maps go into the MEDS, and single-epoch catalogs are used to identify stars for PSF model fitting. 

\subsubsection{PSF estimation}
As discussed in Section~\ref{sec:wl:formalism}, accurate shear inference hinges upon the successful deconvolution of the observation's PSF and galaxy shape measurements. Before their light passes through the atmosphere and telescope, stars are effectively point sources, so the shape and size of their surface brightness profiles axiomatically define the PSF at that location. Using stars as fixed points, the PSF can be interpolated across the rest of the image. 

\begin{figure}[ht!]
    \centering
    \includegraphics[width=0.65\textwidth]{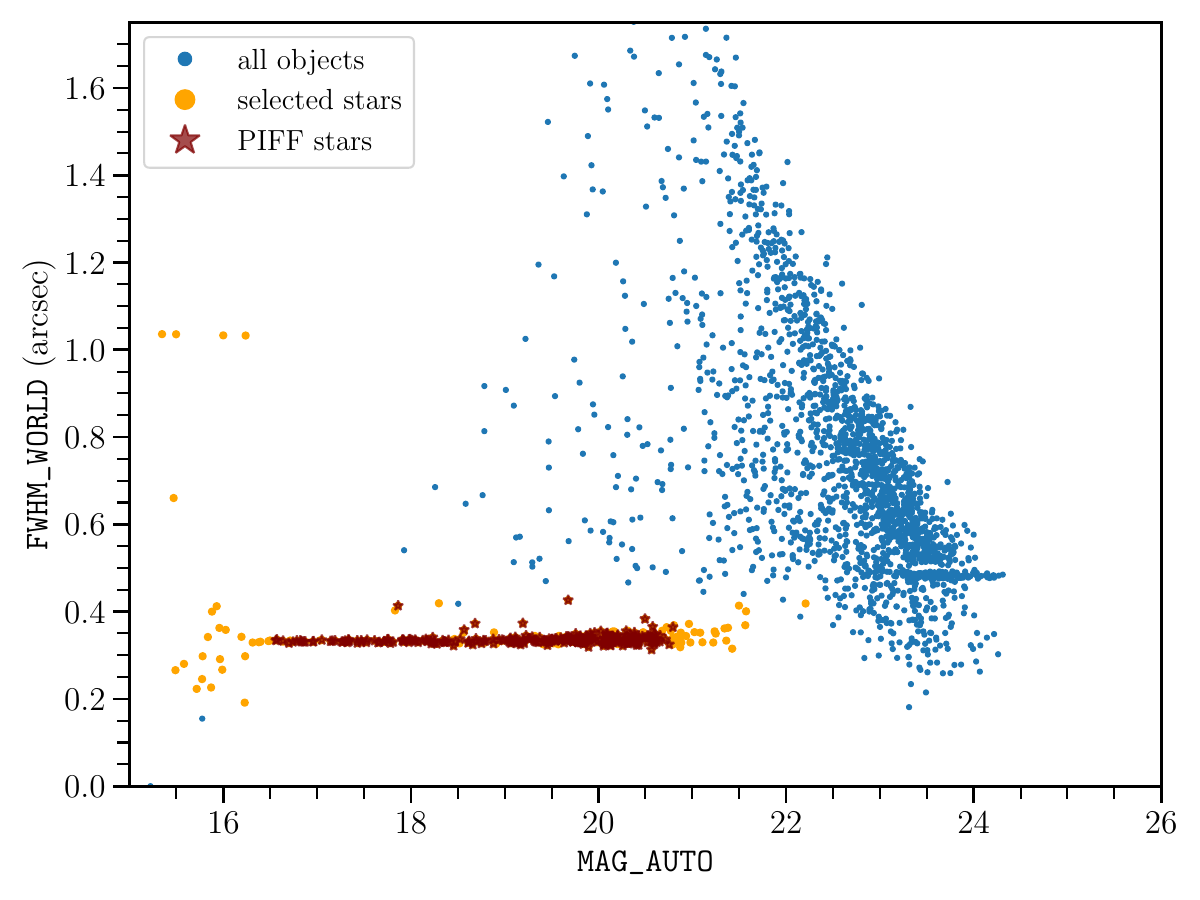}
    \caption{Size-magnitude diagram for objects detected in a single-epoch exposure. Blue points represent all sources; orange points show stars supplied to the PIFF software; dark red star markers show stars selected by the PIFF software for PSF modeling.}
    \label{fig:piffstar_sizemag}
\end{figure}

Star catalogs for PSF modeling are generated with simple selections to the single-exposure detection catalogs based on SExtractor \texttt{CLASS\_STAR}, a minimum signal-to-noise, and a magnitude range. A sample star catalog is highlighted in Figure \ref{fig:piffstar_sizemag}. Should greater sample purity be required, we have incorporated into \texttt{medsmaker} an option to cross-reference candidate stars against a reference catalog and also added the capability to query the Gaia star database~\citep{brown2018Gaia} on the fly. Though the Gaia catalog is relatively shallow, the high purity of the Gaia catalog avoids the problem of star-galaxy confusion. The use of the Gaia catalog for PSF fitting is also considered in \citet{2011ASPC..442..435B}. 

We model \superbit PSFs with the recently introduced \textsc{PIFF} software package\footnote{\url{https://rmjarvis.github.io/Piff/\_build/html/overview.html}}. 
Like most PSF fitters, \textsc{PIFF} takes an input catalog of stars, fits their surface brightness profiles with a user-specified model, interpolates the PSF parameters across the FOV following some schema, and saves the resulting description of the observation's PSF to file. A notable feature of \textsc{PIFF} is that PSF models are expressed in \textit{sky} coordinates, as opposed to the pixel coordinates commonly used in other PSF modeling software. Because high-frequency components of the PSF, e.g., astrometric distortion, vary more smoothly across the detector FOV when considered in sky coordinates, \textsc{PIFF} avoids the ``size bias'' (mismatch between the real and model PSF size) that can affect other PSF fitting software~\citep{2021ascl.soft02024J}. 

Following the DES Y3 approach, we use the \texttt{PixelGrid} model, which treats the PSF profile as a two-dimensional grid of points smoothed by a Lanczos kernel with $n=3$. The total number of free model parameters is then equal to the number of pixels in the grid. 
We also follow the DES Y3 approach to interpolate the PSF model across the FOV by using the \texttt{BasisPolynomial} scheme, which solves for the \texttt{PixelGrid} model parameters (pixel fluxes) in terms of the interpolation coefficients. 

PSF model residuals are quantified with the $\rho$ statistics introduced in \citet{2010MNRAS.404..350R} and expanded in \citet{2016MNRAS.460.1399V}. The $\rho$ statistics below summarize the spatial correlations of size and ellipticity residuals between the real (star) and model PSFs; large values imply a systematic error in the model.  

\begin{align}
        \rho_1(\theta) &\equiv 
            \left\langle 
            \delta e^*_{\rm PSF}(\bm{x}) 
            \delta e_{\rm PSF}(\bm{x}+\bm{\theta})
            \right\rangle\\ 
        \rho_2(\theta) &\equiv 
            \left\langle 
            e^*_{\rm PSF}(\bm{x}) 
            \delta e_{\rm PSF}(\bm{x}+\bm{\theta})
            \right\rangle\\
        \rho_3(\theta) &\equiv 
            \left\langle 
            \left( e^*_{\rm PSF}\frac{\delta T_{\rm PSF}}{T_{\rm PSF}}\right)(\bm{x})
            \left(e_{\rm PSF}\frac{\delta T_{\rm PSF}}{T_{\rm PSF}}\right)(\bm{x}+\bm{\theta})
            \right\rangle\\
        \rho_4(\theta) &\equiv 
            \left\langle 
            \delta e^*_{\rm PSF}(\bm{x})
            \left(e_{\rm PSF}\frac{\delta T_{\rm PSF}}{T_{\rm PSF}}\right)(\bm{x}+\bm{\theta})
            \right\rangle\\  
        \rho_5(\theta) &\equiv 
            \left\langle 
            e^*_{\rm PSF}(\bm{x})
            \left(e_{\rm PSF}\frac{\delta T_{\rm PSF}}{T_{\rm PSF}}\right)(\bm{x}+\bm{\theta})
            \right\rangle
\end{align}
    
\begin{figure}
    \centering
    \includegraphics[width=0.8\textwidth]{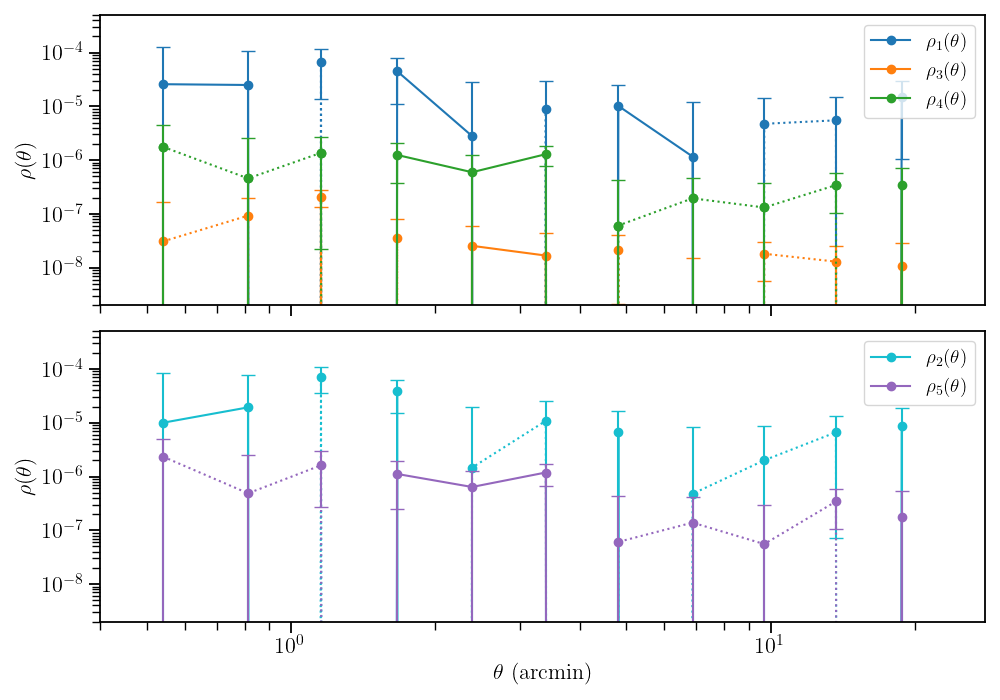}
    \caption{Example of $\rho$ statistics computed from PSF model residuals for a simulated single-epoch exposure. Negative correlations are shown in absolute value and connected to neighboring bins by dotted lines. Isolated points show correlations that changed from positive to negative or vice versa.}
    \label{fig:rho_stats}
\end{figure}

Here $e_{\rm PSF}$ is the ellipticity of the real PSF, i.e., the star ellipticity; $T_{\rm PSF}$ is the size of the real PSF; $\delta e_{\rm PSF}$ is the difference between the ellipticity of the real and model PSFs at position $\bm{x}$; and $\delta T_{\rm PSF}$ is the difference between the sizes of the real and model PSFs at position $\bm{x}$. Brackets denote averages over all pairs within a separation $\bm{\theta}$, and asterisks denote complex conjugates. 
An example of $\rho$ statistics plotted as a function of distance between neighboring stars is shown in Figure \ref{fig:rho_stats}. 

We will also compute the two-point spatial correlations of star and galaxy ellipticities:

\begin{equation}
C_{i} = \langle e_i(\bm{x}) \times e_i( \bm{x}+\bm{\theta})\rangle,\ \ 
    i=\{1,2\}
\end{equation}

where $e_i$ is the \textit{i}th ellipticity component of a PSF-corrected star or galaxy at position $\bm{x}$. The correlations $C_{1/2}$ of galaxy-galaxy pairs should have a relatively high amplitude, reflecting the correlated shear introduced by the galaxy cluster. However, the $C_{1/2}$ functions should vanish when evaluated over star-galaxy pairs, as there should be no correlation between the shapes of circularized stars and PSF-corrected galaxy shapes \citep{mccleary2015mass, 2020ApJ...893....8M}.

\subsubsection{Multi-epoch data structure}

With object cataloging and PSF modeling complete, an instance of the MEDS class is created. For each object in the detection catalog, an entry in the MEDS is made to hold a binary table with postage stamp cutouts from the single-epoch exposures, a PSF model rendering, weights, masks, and segmentation maps. The MEDS is also populated with objects' celestial and image coordinates, original catalog ID number, and WCS information.

\subsection{Galaxy shape measurement}\label{sec:pipeline:metacal}

We measure galaxy ellipticities using the NGMix\footnote{\url{https://github.com/esheldon/ngmix}}
package, which implements Gaussian mixture models to recover the shear from 2D images with good accuracy for even very low S/N galaxies. 
Rather than a single-point estimate, NGMix returns an estimator of the shape from an ensemble of measurements of the galaxy -- generally every epoch, in every filter in which the galaxy was observed. We implement NGMix with a Python wrapper script that creates an instance of the \textsc{NGMixMEDS} class and populates it with observation information for all sources in the supplied MEDS file. 

Because of the high source density of \superbit observations, many of the postage stamps in the MEDS contain not one but two sources: the galaxy of interest and an interloping star or galaxy. Left unmasked, the presence of interlopers introduces a large scatter in the final tangential shear measurements, as NGMix treats both sources as a single galaxy. Following the solution used in DES SV and Y1, we mitigate interlopers using so-called \"{u}berseg masks. These masks are generated using the detection (coadd) image's SExtractor segmentation maps, projected onto the plane of single-epoch exposures. Pixels in the MEDS weight cutouts are set to zero if they are more closely associated with an interloping object than with the galaxy of interest \citep{Jarvis_2016}. 
 
\subsection{Weak lensing shear profile calculation}\label{sec:pipeline:shear_profile}

Tangential and cross shear profiles of galaxy clusters are produced in the \texttt{shear\_profile} module of our pipeline. At this stage, redshift information is added to the galaxy shape fit catalog, selection cuts (including redshift selection) are applied, and Metacalibration responsivities are calculated and applied to galaxy shapes. 
The $(g_{\tan},\, g_{\times})$ shears are computed from $(g_1,\, g_2)$ and then averaged in radial bins from the cluster center. 
Further detail is provided below. 

\subsubsection{Creation of galaxy shear catalog}\label{subsubsec:wl_cuts}
 
A top-level catalog with galaxy shape parameters, responsivity components, detection parameters, and redshifts is generated by joining the SExtractor and NGMix catalogs on sky coordinates $(\alpha, \delta)$ and then matching to a third catalog with redshift information. The galaxy shear catalog is then built from galaxies meeting the following criteria:

\begin{itemize}
    \item 10 $<$ S/N $<$ 1000, where the signal-to-noise measure is the galaxy \texttt{s2n} from NGMix fits
    \item Galaxy size (really, area) 0 $<$ \texttt{T} $<$ 10, where \texttt{T} is in units of arcsec$^{2}$
    \item Ratio of galaxy to PSF size (\texttt{T}/\texttt{Tpsf}) $>$ 1.0 
    \item Galaxy redshift $z_{gal}$ greater than the cluster redshift $z_{cl}$
\end{itemize}
When appropriate, i.e., for a nearly round PSF, we base size and signal-to-noise cuts on the ``roundified'' size and signal-to-noise \texttt{T\_r} and \texttt{s2n\_r}. These selections are based on those in the DES analyses \citep{Jarvis_2016,zuntz2018dark, gatti2021dark}. 

Shear and selection responsivities are calculated from the NGMix \texttt{1p/1m/2p/2m} shape fit parameters to produce responsivity-corrected galaxy shear $(g_1,g_2)$. Selection of background galaxies through redshift cuts is included within the calculation of the selection responsivity. Galaxies are weighted by their shape fit covariances $\sigma^2_{g1/2}$ and a shape noise of $\sigma_{SN} = 0.26$ is based on our own fits to COSMOS galaxies:
\begin{equation}
    \frac{1}{\sigma^2_{SN}+\sigma^2_{g1}+\sigma^2_{g2}}
\end{equation} 
 
\subsubsection{Shear profile calculation}

Response-corrected ($g_1,\, g_2$) moments are transformed into tangential and cross ellipticities $(g_{\tan},\,g_x)$ using the galaxy image coordinates $(x_i,\, y_i)$ and user-specified location of the galaxy cluster center $(x_c,\,y_c)$.

\begin{equation}
    g_{\tan} = -(g_1 \cos(2\phi) + g_2 \sin(2\phi)),
\end{equation}
\begin{equation}\label{eqn:gcross}
    g_{\times} = g_1 \sin(2\phi) - g_2 \cos(2\phi),
\end{equation}
\begin{equation}\label{eqn:phi}
    \phi = \frac{1}{2}\arctan\left(\frac{y - y_c}{x - x_c}\right)
\end{equation}

\begin{figure}
   \centering
   \includegraphics[width=\textwidth]{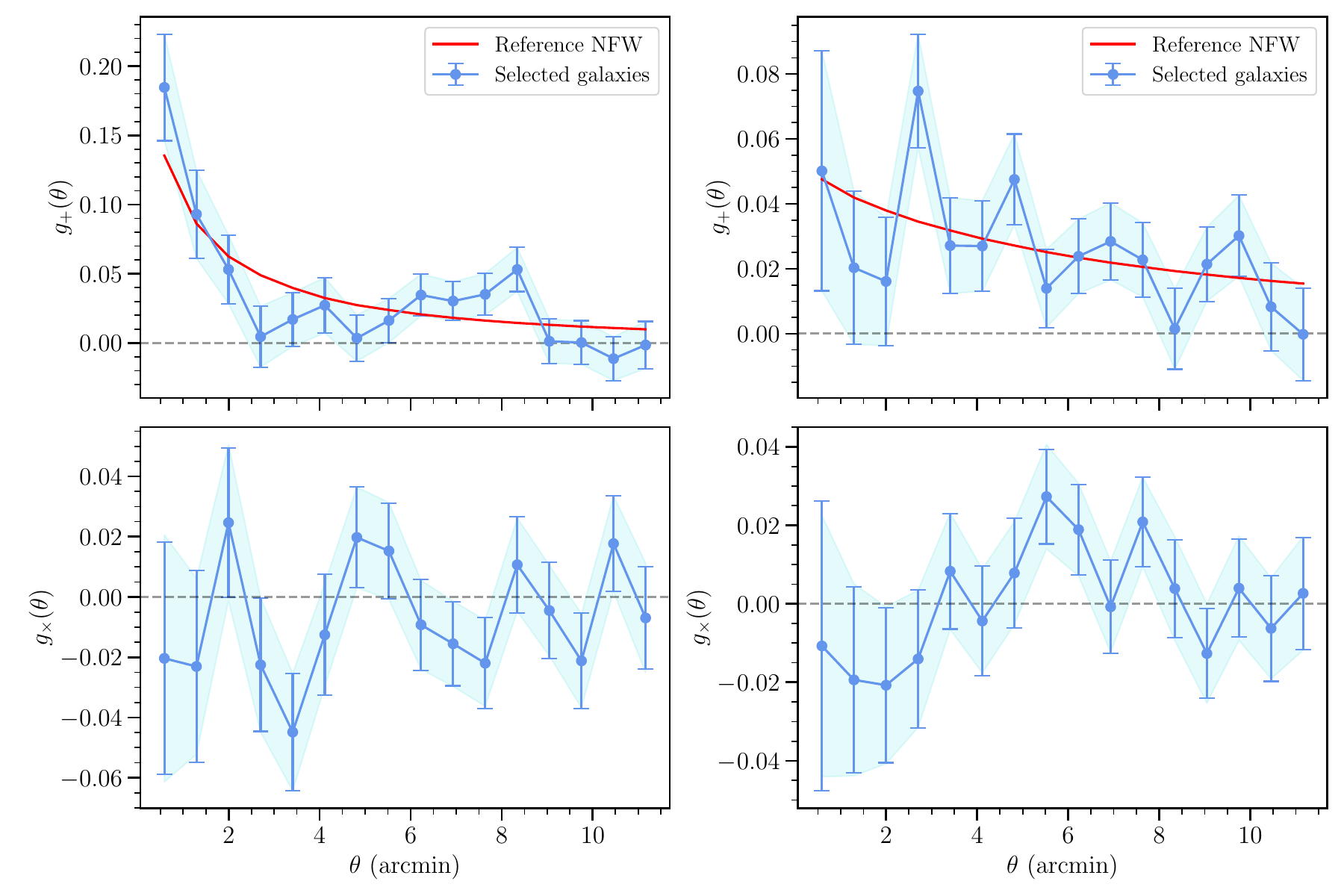}   
   \caption{Examples of tangential (top panels) and cross shear profiles (bottom panels) for single galaxy cluster realizations in $lum$. Left: Realization of a cluster with $M_{200c}=4.1\ee{14}\, M_{\odot} \, h^{-1}$ and $z=0.45$. Right:  Realization of a cluster with $M_{200c}=4.1\ee{14}\, M_{\odot}\, h^{-1}$ and $z=0.059$. The selections described in Section \ref{subsubsec:wl_cuts} were applied, yielding 31.3 and 31.6 galaxies arcmin$^{-2}$ for the $z=0.059$ and $z=0.3$ clusters, respectively.}
   \label{fig:MiscShearProfiles1}
\end{figure}

The \texttt{DescrStatsW} class of the Python package \texttt{statsmodels} package to compute weighted averages of $(g_{\tan},\, g_{\times})$ in radial bins about the cluster center. The final outputs are a shear profile catalog with averaged $(g_{\tan},\, g_{\times})$ and a plot of the cluster's cross and tangential shear profiles. Two examples from simulated observations are shown in Figure \ref{fig:MiscShearProfiles1}. 

\subsection{Shear bias estimator}

While we do not attempt shear calibration in this analysis, we have developed an estimator for shear bias tailored to cluster shear tangential profiles. It is included in the pipeline to support future efforts. 

Borrowing the language of large cosmological surveys, we express the difference between input (simulated) and output (measured) tangential shears as a \textit{shear bias} $\alpha$, which we quantify with a maximum likelihood estimator $\hat{\alpha}$. Each galaxy's measured shear $g_{\tan}$ is considered a random sample of the true halo shear $g_{\rm true}$ at the galaxy's position. The joint probability distribution (likelihood) $\mathcal{L}$ of the data then follows a multivariate Gaussian in which the mean shear $\langle g_{\rm tan}\rangle$ converges to $\alpha\,g_{\rm true}$ in a radial bin, where 

\begin{equation}
    \alpha = \frac{\langle g_{\tan}\rangle}{\langle g_{\rm true}\rangle}
\end{equation}

If the measurements $g_{\rm tan}$ are unbiased measurements of the true shear $g_{\rm true}$, then $\alpha \equiv 1$ and $\langle g_{\rm tan}\rangle = g_{\rm true}$. Any $\alpha \neq 1$ indicates a biased measurement.

To obtain an optimal estimator $\hat{\alpha}$ for the tangential shear bias $\alpha$, we express the log-likelihood as
\begin{equation}
\log \mathcal{L} = - \rm \left [\mathbf{g}_{\tan} - \alpha \, \mathbf{g}_{\rm true}\right ] \frac{\mathbf{C}^{-1}}{2} \left[\mathbf{g}_{\tan} - \alpha \, \mathbf{g}_{\rm true}\right ]^{\rm T}
\label{eq:likelihood}
\end{equation}
The measurement uncertainties of the data $\mathbf{g}_{\tan}$ are expressed as the covariance $\mathbf{C}$. Note that matrix quantities are written in boldface. 
Differentiating Equation \ref{eq:likelihood} with respect to $\alpha$, setting the result to zero, and then solving for $\alpha$, we obtain the maximum likelihood estimator for shear bias: 

\begin{equation}
\hat{\alpha} = \frac{\mathbf{g}_{\rm true}^{\rm T}\ \mathbf{C}^{-1} \  \mathbf{g}_{\tan}} {\mathbf{g}_{\rm true}^{\rm T} \  \mathbf{C}^{-1 } \ \mathbf{g}_{\rm true}}
\label{eq:estimator}
\end{equation}

The uncertainty on $\hat{\alpha}$ is given by the the Cram\'{e}r-Rao bound:
\begin{equation}
\sigma^2_{\hat{\alpha}} = \frac{1}{\mathbf{g}_{\rm true}^{\rm T} \  \mathbf{C}^{-1 } \ \mathbf{g}_{\rm true}}
\label{eq:CRO}
\end{equation}

\begin{figure}
\begin{center}
\includegraphics[width=0.7\textwidth]{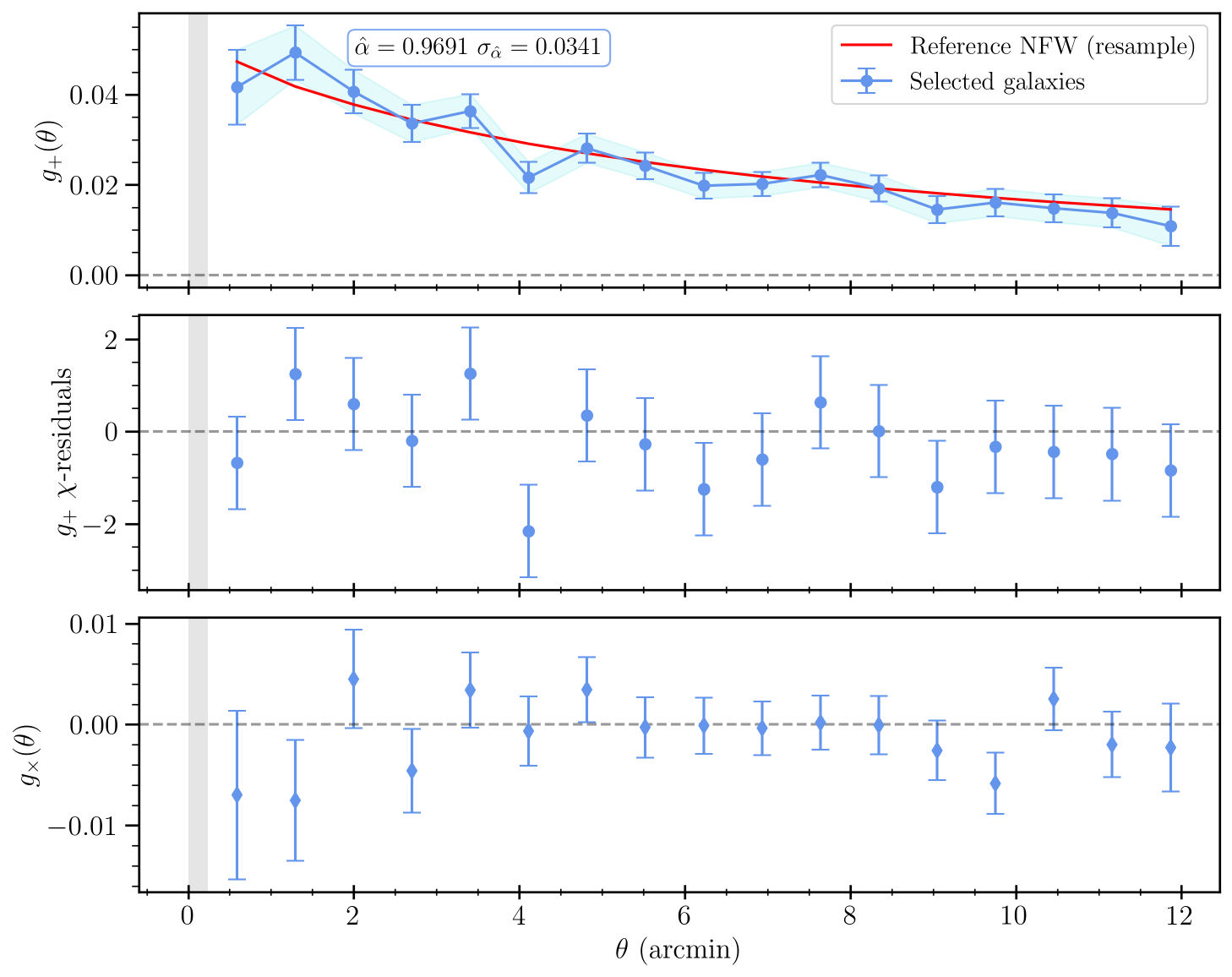}\\

\caption{Example of the setup for shear bias ($\alpha$) estimation. Top panel: mean tangential shear. Blue points are the mean galaxy shears in a radial bin for 30 realizations of $z=0.059$ clusters; error bars are standard errors of the mean. The input NFW tangential shear is plotted as a red line. Middle panel: difference between measured and input tangential shears. Bottom panel: cross shear. The gray-shaded regions indicate the regime where the linearized implementation of Metacalibration is invalid. The \superbit shear calibration analysis will be presented in a forthcoming paper by S. Everett et al.}
\label{fig:AnalysisMeanShearAlpha}
\end{center}
\end{figure}

An unbiased cluster tangential shear measurement has $\hat{\alpha} = 1 \, \pm \, \sigma_{\hat{\alpha}}$. The goal for \superbit shear calibration will be a shear bias consistent with unity within the mass uncertainty of the full cluster sample (2--3 \%). As the value of $\hat{\alpha}$ calculated from a large number of simulations is a useful metric for shear calibration, the pipeline contains tools for the calculation of the average $\hat{\alpha}$ as well. An example setup is shown in Figure \ref{fig:AnalysisMeanShearAlpha}. The \superbit shear calibration analysis will be presented in S. Everett et al. (in preparation). 

\section{Simulated galaxy cluster observations}\label{sec:the_simulations}


\begin{figure}
   \centering
   \includegraphics[width=0.75\textwidth]{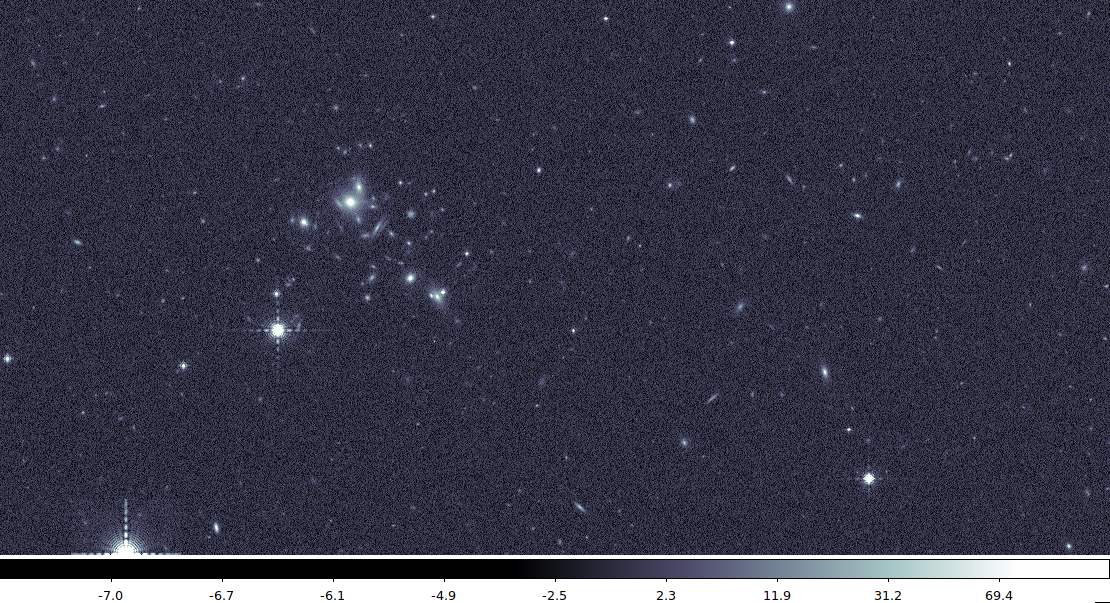} 
   \caption{Simulated three-hour \superbit observation of a galaxy cluster in the $b$ band using the jitter+optics (`optics-on') PSF. The central $6\arcmin \times 3\arcmin$ region of the simulation is shown. Pixel values represent background-subtracted ADU.}
   \label{fig:sim}
\end{figure}

To plan observations and calibrate the analysis pipeline for \superbit's science flight, we have used GalSim \citep{2015A&C....10..121R} to produce mock observations of galaxy clusters. These simulate 3~hour observations in each of \superbit's \umag, \bmag, \lum and \shape filters, divided into \texttt{n\_exp}=36 individual, dithered exposures of \texttt{exp\_time}=$300$~seconds. The central region of a full 3-hour observation of one simulated cluster is shown in Figure \ref{fig:sim}. For each cluster, we create 30 mock sets of images with independent distributions of 
stars, cluster member galaxies, and field galaxies both in front of and behind the galaxy cluster. We store a truth catalog containing the objects' positions, sizes, fluxes, redshifts, and applied lensing distortion (for galaxies behind the cluster). For distortion calculations, we set $\Omega_M = 0.3$ and $\Omega_\Lambda = 0.7$. 

Simulated clusters have mass $M_{200c} = 4.1 \ee{14}\, M_{\odot} \, h^{-1}$ (the mean mass of clusters in the \superbit target list) and three redshifts ($z=0.059, 0.3, 0.45$).
Cluster mass distributions are modeled with Navarro, Frenk, and White (1996; NFW) density profiles:
\begin{equation}
\rho(r) = \frac{\rho_0}{
	\frac{r}{R_S}\left(
	1 + \frac{r}{R_S}
	\right)^2
	}
\end{equation}

\subsection{Point spread function and stars}

\begin{figure}
   \centering
   \includegraphics[width=\textwidth]{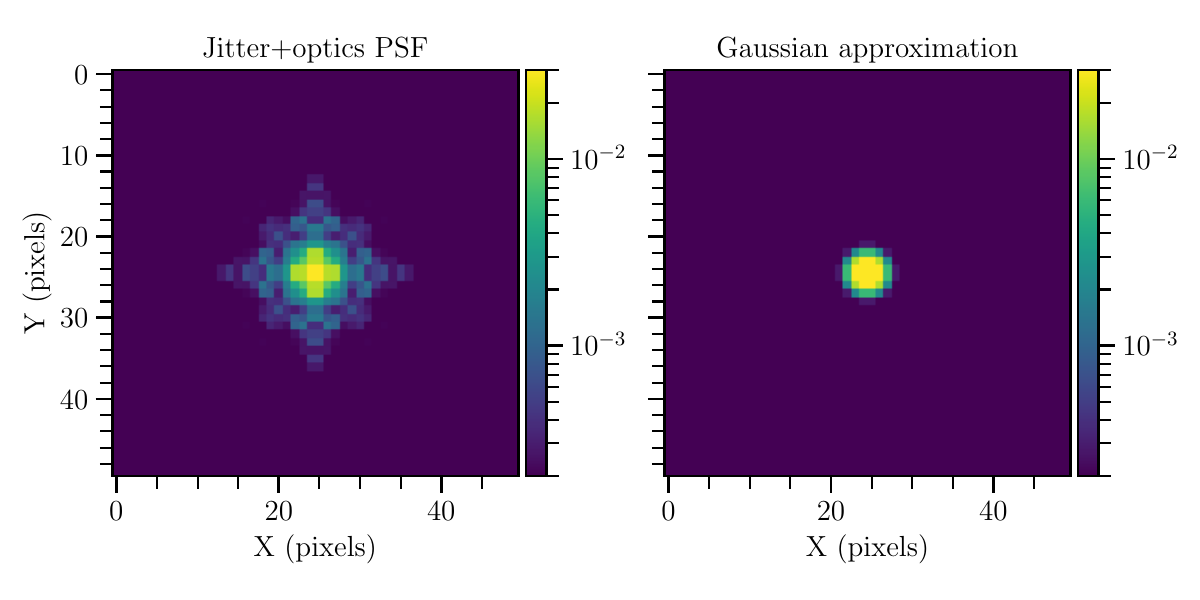} 
   \caption{Left: rendering of \superbit effective (pixel-convolved) PSF in \bmag filter. Right: Gaussian approximation to \bmag PSF with FWHM = $0.315\arcsec$. In both panels, intensity values have been normalized so that the total flux in each image sums to one; the color scale represents relative flux or intensity.}
   \label{fig:psf}
\end{figure}

The \superbit PSF is well modeled with two components (jitter+optics or `optics-on') that combine the residual telescope jitter measured during test flights with spherical aberrations for the \superbit optical train derived with ray-tracing software. 
However, we base survey forecasts on an (`optics-off') Gaussian approximation to the PSF because of a temporary limitation in the NGMix method that we use for shape measurement. NGMix
does not currently include templates for diffraction-limited PSFs, and tends to over-estimate the PSF size $T_{\rm PSF}$ by about $\sim 50\%$. Given that almost all weak lensing analyses select galaxies based on their size relative to PSF size ($T/T_{\rm PSF}$), this artificially decreases the source density. An extension to the NGmix template set will be presented in the shear calibration paper by S.\ Everett et al. Meanwhile, we implement Gaussian approximations to the jitter+optics PSF, with FWHM of  $0.278\arcsec$ in \umag, $0.315\arcsec$ in \bmag, $0.333 \arcsec$ in \lum, and $0.37\arcsec$ in \shape. These values are the combination of the jitter FWHM of $0.05\arcsec$ and the FWHM obtained with ray-tracing models of the \superbit optical train in each bandpass. The `optics-on' and `optics-off' versions of the $b$-band PSF are compared in Figure \ref{fig:psf}.

We simulate the spatial clustering and magnitude distribution of foreground stars by sampling Gaia DR2 catalogs \citep{2018yCat.1345....0G} at the RA and Dec coordinates of 52 of \superbit's target clusters. We convert Gaia G/G$_{\rm BP}$ fluxes to \superbit AB fluxes, then choose one of these star fields at random for each realization of mock images. Because the star fields span a range of galactic latitudes, this effectively marginalizes over stellar number density when predicting shear biases. 

\subsection{Galaxies}

Our simulation input catalog is a hybrid of two different COSMOS catalogs. A full description for generating the mock \superbit source galaxy catalog will appear in a paper by A.\ Gill et al. (in preparation); a high-level overview is presented here. 

The baseline is the UltraVISTA-DR2 region of the COSMOS 2015 catalog~\citep{2016ApJS..224...24L}, which contains 518,404 galaxies with high-quality redshifts spread out over 1.5 deg$^2$. The number density, redshift, and magnitude distributions of our simulated background source galaxies are drawn directly from COSMOS 2015. To convert COSMOS 2015 fluxes to their equivalent in \superbit bandpasses, we access the spectral energy distribution fits from the EL-COSMOS project \citep{saito}, convolve these with the wavelength-dependent OTA throughput, detector QE, and filter transmission curves, and finally integrate counts over the collecting area of the \superbit mirror (cf.~Section \ref{sec:superbit:instrument}). 

We add morphological information to COSMOS 2015 with a heuristic match in luminosity ($m_{\rm C15}$) and redshift ($z_{\rm C15}$) to galaxies in the GalSim COSMOS F814W$<$25.2 catalog. In our simulations, galaxies are drawn as single-component Sers{\'i}c profiles with half-light radius $R_{1/2}$ and index $n$, position angle $\phi$, and major-to-minor axis ratio $q$. 

Parameter values are chosen with the following algorithm:
    \begin{itemize}
        \item In the best-case scenario of $z_{\rm C15} < 5$ and 
        $18 <m_{\rm C15}<25.2$, a source is selected from the GalSim COSMOS that best matches the COSMOS 2015 galaxy and its shape parameters are assigned to the COSMOS 2015 galaxy.
        \item If $z_{\rm C15} < 5$ and 
        $25.2 <m_{\rm C15}<30$, a source is selected from GalSim COSMOS that best matches the COSMOS 2015 galaxy redshift and used to set the half-light radius. The Ser{\'i}c index $n$ is selected from a uniform distribution $U[0,\,4]$. The position angle $\phi$ is also chosen from a uniform distribution $U[-2,\,2]$ radians. The axis ratio $q$ is selected from a uniform distribution $U[0.1,\,1]$.
        \item If $z_{\rm C15} > 5$ but 
        $18 <m_{\rm C15}<25.2$, $n$, $q$ and $\phi$ are chosen based on the closest match in $m_{\rm F814W}$ between GalSim COSMOS and COSMOS 2015. The half-light radius $R_{1/2}$ is randomly chosen from a uniform distribution $U = [5,20]$ pixels (plate scale = $0\farcs03$/pixel).
        \item All other $z_{\rm C15}$ and $m_{\rm C15}$ cases correspond to outliers with no equivalents in the GalSim COSMOS catalog. In this instance, all galaxy shape parameters are chosen from uniform distributions. 
        
   \end{itemize}

While GalSim does have ready-made galaxy catalogs available, their maximum depth of F814W = 25.2 would limit our ability to simulate deep \superbit observations. Moreover, the number of galaxies with photometric redshifts has increased since 2007 (the year of the original GalSim COSMOS catalog's release). These limitations motivated us to create our own galaxy catalog for simulations.

\subsection{Simulation procedure} 

First, we initialize the random number generators for stars, source galaxies, cluster galaxies, noise and dither offsets, passing any seeds set in the GalSim configuration file.

The blank exposure is represented with an instance of the GalSim object (GSObject) \texttt{ImageF} set to match the \superbit instrument properties of Section~\ref{sec:superbit:instrument}, and includes a model world coordinate system (WCS). The image is filled with the raw sky background derived in \cite{gill2020optical}; approximately 45 ADU for a 300-second exposure in the \bmag filter. 

The cluster lensing potential is represented with an instance of the \texttt{NFWHalo} class. The halo concentration is set to 4 in all simulations. 

For each source galaxy to be injected into the image, the following process is repeated. A galaxy entry is randomly drawn from the \superbit mock galaxy catalog and assigned some right ascension and declination on the observation. The galaxy's photometric redshift, shape parameters, and flux in the \superbit filter of choice are accessed from our mock galaxy catalog. The galaxy image is created as an instance of \texttt{Sersic} with shape parameters set to the catalog values. To convert the catalog flux from units of photoelectrons s$^{-1}$ to equivalent observed analog-to-digital units (ADU), we multiply the flux by the exposure time and the gain. 

The source galaxy object is sheared and magnified according to its redshift with the \texttt{NFWHalo} object, or if the source galaxy redshift is below the cluster redshift, the galaxy's magnification and distortion are set to 1 and 0 respectively. The galaxy image then convolved with the PSF model. For later reference, the galaxy position, lensing magnification, reduced shear moments, redshift, and stamp flux are passed to a truth catalog. Finally, the galaxy image is converted to a ``stamp'' GSObject and drawn onto the observation at the appropriate coordinates. We inject a fiducial number of 99 galaxies per square arcminute.

Cluster galaxies are generated in much the same way as source galaxies, except that they are concentrated in the center of the observation and no lensing distortion is applied. The number of cluster galaxies (30) is set to approximately match the source density of bright cluster galaxies in the 2019 Abell 2218 observation. They are uniformly distributed in a circle of radius 200 pixels ($28\arcsec$). A random offset is added of $\pm 50$ pixels, about $7\arcsec$ per galaxy. Because the cluster galaxies are generally large and bright, the default GalSim COSMOS F814W $<$ 23.5 sample catalog is sufficient for modeling cluster galaxy sizes and brightnesses. For recording in the truth catalog, they are assigned a redshift equal to the redshift in the \texttt{NFWHalo} class. 

We have an ensemble of catalogs containing star positions and brightness. These catalogs are made using the Gaia satellite observations of the galaxy clusters in \superbit's planned target list. For each simulation, we select a catalog and draw the same number of stars as observed by Gaia, using their fluxes to accurately represent the stars' brightness, while the spatial density is also preserved. Star positions, however, are not specifically replicated.

Pre-seeing stars are modeled as \texttt{DeltaFunction} objects, with a flux randomly drawn from the selected cluster's Gaia catalog of real stars. The star model is convolved with the same PSF model as above, before itself being drawn into the observation. Unless otherwise specified in the configuration file, the total number of stars injected over the entire field of view matches the number of entries in the selected Gaia catalog.

Once injection of all stars and background and cluster galaxies is complete, we add dark current to the image. The final step is application of the \texttt{CCDNoise} method, which adds Poisson noise to the image based on the pixel values (including read noise). At this point, the simulated observation may be saved to file, and the process is repeated up to the total desired exposure time, in each desired filter.

\begin{figure}
    \centering
    \includegraphics[width=0.7\textwidth]{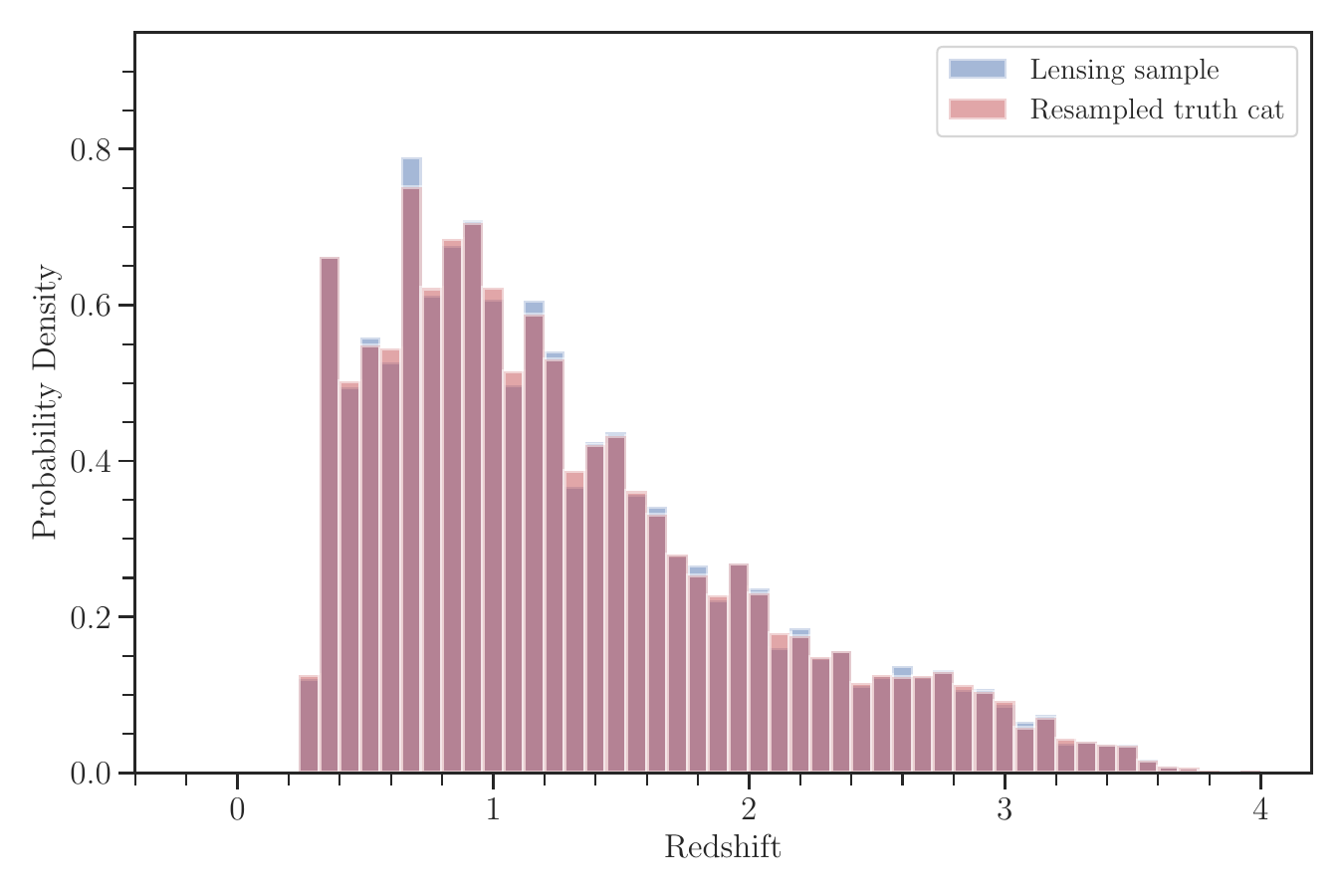}
    \caption{Redshift distribution for weak lensing source galaxies and bootstrapped reference 
    NFW catalog in one simulated observation of a cluster at $z=0.3$.} 
    \label{fig:nfw_resampled}
\end{figure}

To provide a reference for shear bias calculations, NFW tangential shear catalogs are generated in every $(M, z)$ bin with a modified version of the simulations code. The redshift distributions of the reference NFW catalogs are identical to the input COSMOS catalog; however, they will differ significantly from the redshift distributions of the final mock observation catalogs. We circumvent this problem by resampling the NFW references catalog with a Monte Carlo rejection sampling algorithm until the redshift distributions match the mock observation catalogs. Figure~\ref{fig:nfw_resampled} shows an example of the resulting, nearly indistinguishable redshift distributions. 


\section{Results}\label{sec:results}
    

Having developed this data analysis infrastructure, we now consider its application to our simulated galaxy cluster observations. Table \ref{tab:summary} summarizes the mean source density, imaging depth, and galaxy redshift distributions for mock \superbit observations of clusters in three redshift bins: $z=0.059$, $z=0.3$, and $z=0.45$. These results are computed from 30 independent realizations in each redshift bin for a total of 90 unique cluster fields. 

We estimate survey properties for the total number of galaxies observed (``all galaxies'' in Table \ref{tab:summary}) and lensing-analysis galaxies that pass selection cuts in Section \ref{subsubsec:wl_cuts} (``lensing''). To separate the effect of redshift cuts from the rest of the lensing selections in \ref{subsubsec:wl_cuts}, we also compute survey properties for the background galaxies for each cluster (``$z_{gal} > z_{clust}$'') without any size selections. All quantities are computed on the coadded images and obey the following color convention in plots: \umag is shown in pink, \bmag in blue, \lum in orange, and \shape in red. All magnitudes are expressed in the AB magnitude system. 

\begin{deluxetable}{ccccccc}\label{tab:summary}

\tabletypesize{\small}

\tablecaption{Forecast Observation Depths and Redshifts}

\tablehead{\colhead{Cluster $z$} & \colhead{Galaxy sample} & \colhead{Filter} & \colhead{Source density} & \colhead{$S/N=10$ depth} & \colhead{Median $z$} & \colhead{Mean $z$} \\
\colhead{} & \colhead{} & \colhead{} & \colhead{($N_{\rm gals}$ arcmin$^{-2}$)} & \colhead{(AB mag)} & \colhead{} & \colhead{} } 

\startdata
0.059 & All galaxies & u & 15.4 & 25.5 & 0.9 & 1.0 \\
0.059 & All galaxies & b & 43.1 & 26.3 & 1.1 & 1.3 \\
0.059 & All galaxies & lum & 45.5 & 26.3 & 1.1 & 1.3 \\
0.059 & All galaxies & shape & 36.5 & 25.2 & 0.9 & 1.2 \\
\hline
0.059 & $z > z_{\rm clust}$ & u & 15.2 & 25.5 & 0.9 & 1.1 \\
0.059 & $z > z_{\rm clust}$ & b & 42.9 & 26.3 & 1.1 & 1.3 \\
0.059 & $z > z_{\rm clust}$ & lum & 45.2 & 26.3 & 1.1 & 1.3 \\
0.059 & $z > z_{\rm clust}$ & shape & 36.3 & 25.2 & 0.9 & 1.2 \\
\hline
0.059 & Lensing & u & 9.1 & 25.4 & 0.9 & 1.0 \\
0.059 & Lensing & b & 31.4 & 26.3 & 1.0 & 1.2 \\
0.059 & Lensing & lum & 33.5 & 26.2 & 1.0 & 1.2 \\
0.059 & Lensing & shape & 26.0 & 25.1 & 1.0 & 1.2 \\
\hline
0.3 & $z > z_{\rm clust}$ & u & 12.9 & 25.4 & 1.0 & 1.1 \\
0.3 & $z > z_{\rm clust}$ & b & 38.5 & 26.3 & 1.2 & 1.3 \\
0.3 & $z > z_{\rm clust}$ & lum & 40.6 & 26.3 & 1.1 & 1.3 \\
0.3 & $z > z_{\rm clust}$ & shape & 32.1 & 25.2 & 1.0 & 1.3 \\
\hline
0.3 & Lensing & u & 7.6 & 25.4 & 0.9 & 1.1 \\
0.3 & Lensing & b & 28.3 & 26.2 & 1.1 & 1.3 \\
0.3 & Lensing & lum & 30.2 & 26.2 & 1.1 & 1.3 \\
0.3 & Lensing & shape & 23.2 & 25.1 & 1.0 & 1.2 \\
\hline
0.45 & $z > z_{\rm clust}$ & u & 11.1 & 25.5 & 1.1 & 1.2 \\
0.45 & $z > z_{\rm clust}$ & b & 34.5 & 26.3 & 1.2 & 1.4 \\
0.45 & $z > z_{\rm clust}$ & lum & 36.4 & 26.3 & 1.2 & 1.4 \\
0.45 & $z > z_{\rm clust}$ & shape & 28.2 & 25.2 & 1.1 & 1.3 \\
\hline
0.45 & Lensing & u & 6.4 & 25.4 & 1.0 & 1.2 \\
0.45 & Lensing & b & 25.3 & 26.2 & 1.2 & 1.4 \\
0.45 & Lensing & lum & 27.2 & 26.2 & 1.2 & 1.4 \\
0.45 & Lensing & shape & 20.5 & 25.1 & 1.1 & 1.3 \\
\enddata

\tablecomments{Results are based on three hours of integration time per band per cluster. The $z>z_{clust}$  and ``all galaxies'' samples have a \SN{5} selection. ``Lensing'' galaxies pass the selection criteria listed in Section \ref{subsubsec:wl_cuts}.}

\end{deluxetable}

\subsection{Source density}\label{sec:results:source_density}

\begin{figure}
   \centering
   \includegraphics[width=0.75\textwidth]{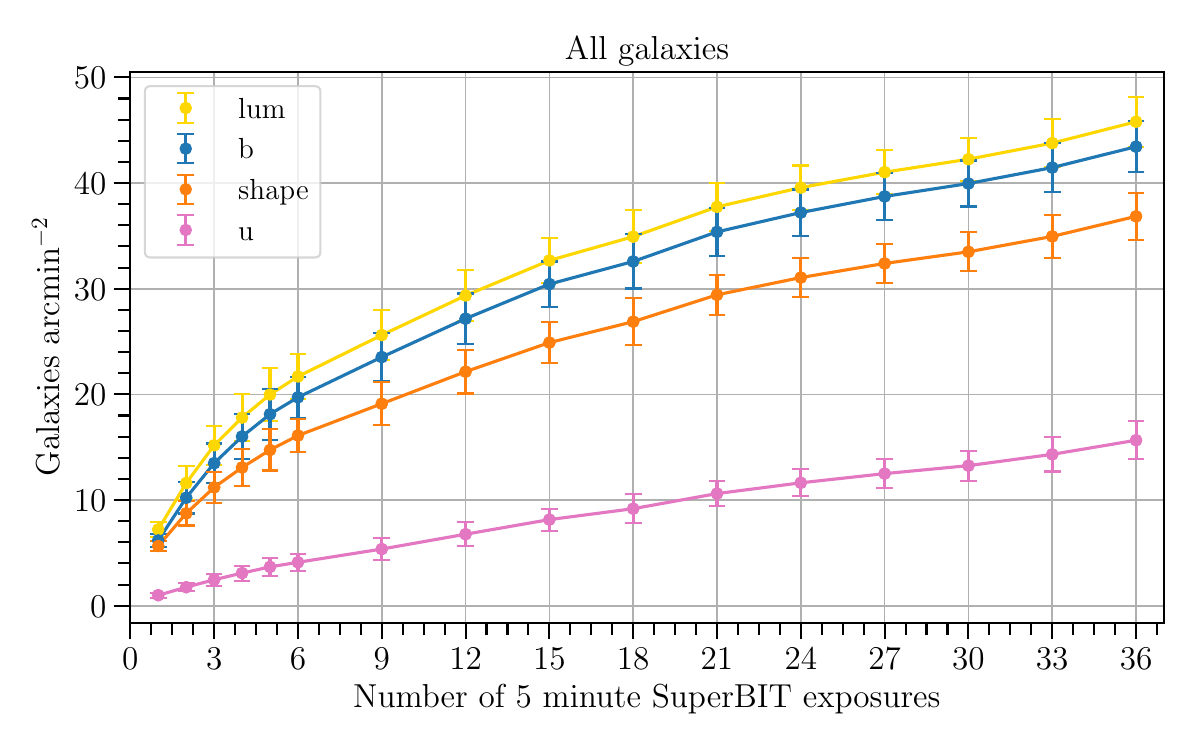} 
   \caption{Galaxy number density as a function of integration time (measured in 5-minute increments) in four \superbit bandpasses. Points are the mean values across 30 simulated galaxy clusters with $M=4.1\ee{14}\, M_{\odot} \, h^{-1},\, z=0.059$; error bars are standard errors of the mean. Beyond a requirement that galaxy \SN{5}, no selections on galaxy redshifts or fit parameters are made.}
   \label{fig:joined_match}
\end{figure}

\begin{figure}
   \centering
   \includegraphics[width=0.75\textwidth]{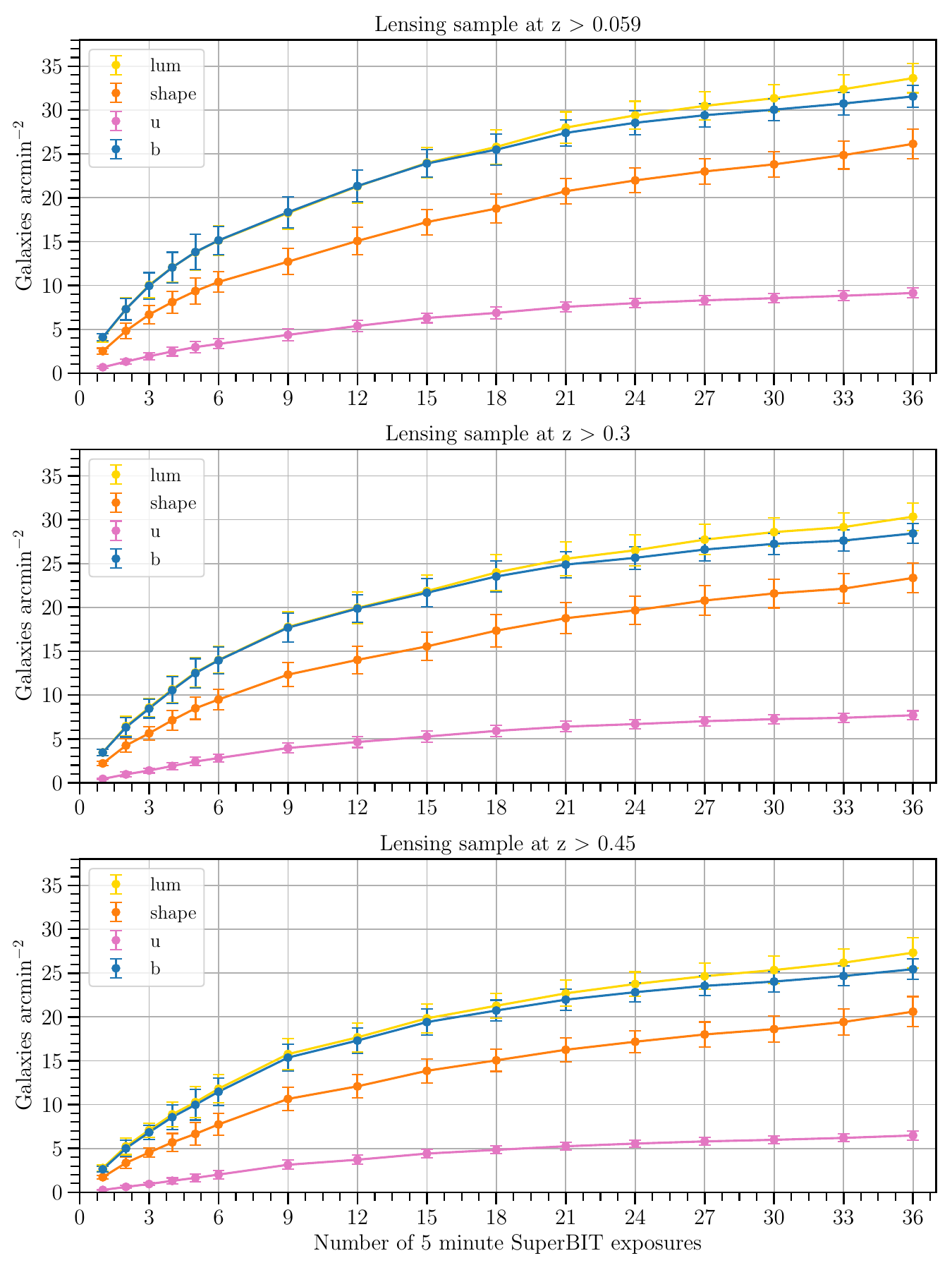} 
   \caption{Background galaxy number density as a function of total
     integration time on clusters at $z=0.059$ (top), $z=0.3$
     (middle), and $z=0.45$ (bottom). Plotted values are the mean of
     30 simulated observations of clusters with $M=4.1\ee{14}$ and the
     indicated redshift; error bars are the standard errors of the
     mean. The source densities reflect the selection criteria for clusters' respective ``lensing'' samples.}
   \label{fig:ann_match}
\end{figure}

We compute the galaxy number (source) density as a function of exposure time as follows. Upon completion of a cluster realization, a script generates a list of exposures that are a subset of the total number. Next, a pared-down version of \texttt{medsmaker} combines the exposures into a coadd and produces a source catalog, which is then matched to the galaxy and lensing analysis catalogs of the full observation. The process is repeated for 1-6 exposures and then intervals of 3 exposures. Once this process is complete for all thirty realizations in that $(M,\, z)$ and bandpass bin, the script computes summary statistics such as mean and standard deviation of galaxy catalog lengths for the $n$-exposure coadds. 

Figures \ref{fig:joined_match} and \ref{fig:ann_match} show mean number of galaxies per square arcminute as a function of on-sky integration time. Results are shown for the $u$, $b$, $lum$, and $shape$ bands. Error bars are standard error of the mean across the 30 cluster realizations of each redshift bin. Integration time is expressed in number of coadded five-minute exposures (the fiducial exposure time) to reach a total of 3 hours ($36 \times 5 minutes$). 

Total galaxy number densities are shown in Figure \ref{fig:joined_match}; these samples have no selections on galaxy shape fits or redshifts beyond a SExtractor \texttt{SNR\_WIN} $>5$ cut. The source densities for three hours of integration time are, 45.5 galaxies per square arcminute in $lum$, 43.1 in $b$, 36.5 in $shape$, and 15.4 in $u$. 

The growth of source density is well fit by a logarithmic function. In the planned shear measurement band $b$, 
\begin{equation}
N_{gals} = 11.01\, \log_2(2.99 + N_{exp}) - 15.34 \label{eqn:sourcedensity}
\end{equation}
Extrapolating outwards, increasing the $b$ source density from 43 \Neff to 50 \Neff would take an additional 1.3 hours of observation. 

The background galaxy number densities in Figure \ref{fig:ann_match} include the lensing sample selections of Section \ref{subsubsec:wl_cuts}. Lensing-analysis samples for clusters at $z=0.059$ have mean source densities of 33.5 \Neff in $lum$; 31.4 \Neff in $b$; 26.0 in $shape$; and 9.1 in $u$ (though we would not attempt weak lensing measurements in $u$). For clusters at $z=0.3$, the corresponding source densities are 30.2 \Neff in $lum$; 28.3 \Neff in $b$; 23.2 in $shape$; and 7.6 in $u$. Source densities clusters at $z=0.45$ (the highest redshift bin considered) have mean source densities of 27.2 \Neff in $lum$; 25.3 \Neff in $b$; 20.5 in $shape$; and 6.4 in $u$. 

To separate the effect of redshift cuts from the rest of the lensing selections in \ref{subsubsec:wl_cuts}, we also calculate source densities of background galaxies with no additional selections. Table \ref{tab:summary} shows that redshift cuts alone produce more modest drops in source density than the lensing selections. The change in source density for a cluster at $z=0.059$ is insignificant within error bars, but lensing selections reduce the source density in $b$ by 27\%, from 43.1 to 31.4 \Neff. The source density behind $z = 0.3$ is 38.5 galaxies arcmin$^{-2}$ (about 10\% drop from 43.1), but the rest of the lensing selections leaves 28 galaxies arcmin$^{-2}$ (two-thirds of the original source density). Similarly, the source density behind $z = 0.45$ is 34.5 galaxies arcmin$^{-2}$ in $b$ (a 20\% drop), while the lensing sample has a source density of 25.3 galaxies arcmin$^{-2}$ (40\% lower than the full galaxy sample). We find that lensing-analysis selections tend to decrease the source densities more strongly than redshift cuts alone.

\subsection{Depths}\label{sec:results:depths}

Depth, or the limiting magnitude for some threshold, is a commonly used figure of merit in astronomical surveys. We adopt the magnitude limit corresponding to a fixed $\sim S/N=10$ threshold ($9.8-10.2$) based on $\delta F / F \sim 0.1$, where $F =$ \texttt{FLUX\_AUTO} and $\delta F =$ \texttt{FLUXERR\_AUTO} \citep{2018ApJS..239...18A}.

\begin{figure}
   \centering
   \includegraphics[width=0.6\textwidth]{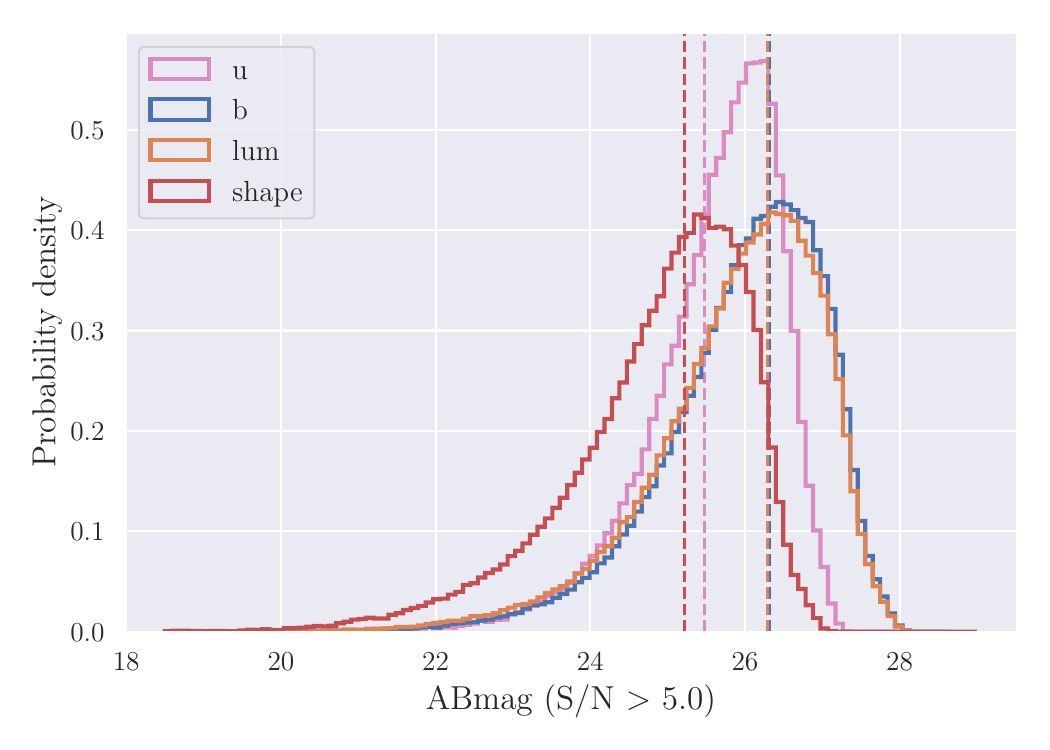}\\
   \includegraphics[width=0.6\textwidth]{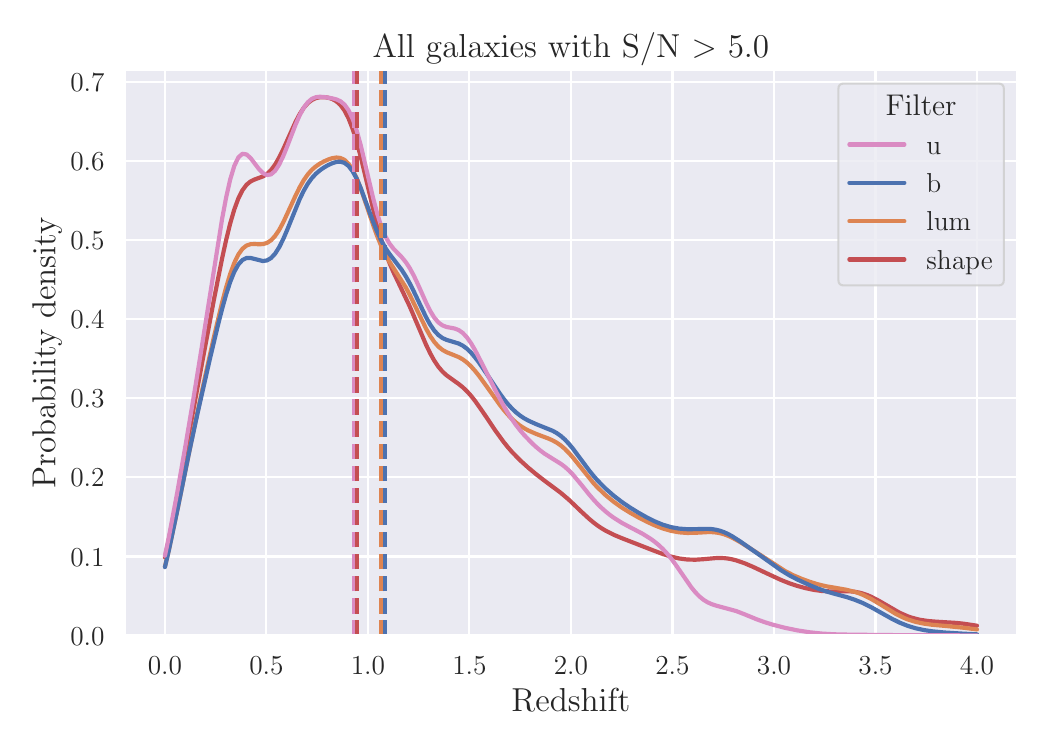}
   \caption{Galaxy brightness histograms (top) and redshift distributions (bottom) for the ``all galaxies'' sample with only a \SN{5} selections. Dotted lines mark the $S/N = 10$ limiting magnitudes (top) and the median redshift (bottom), respectively, in each filter.}
   \label{fig:allgals_lumfunc_zdist}
\end{figure}

\begin{figure}
   \centering
   \includegraphics[width=0.9\textwidth]{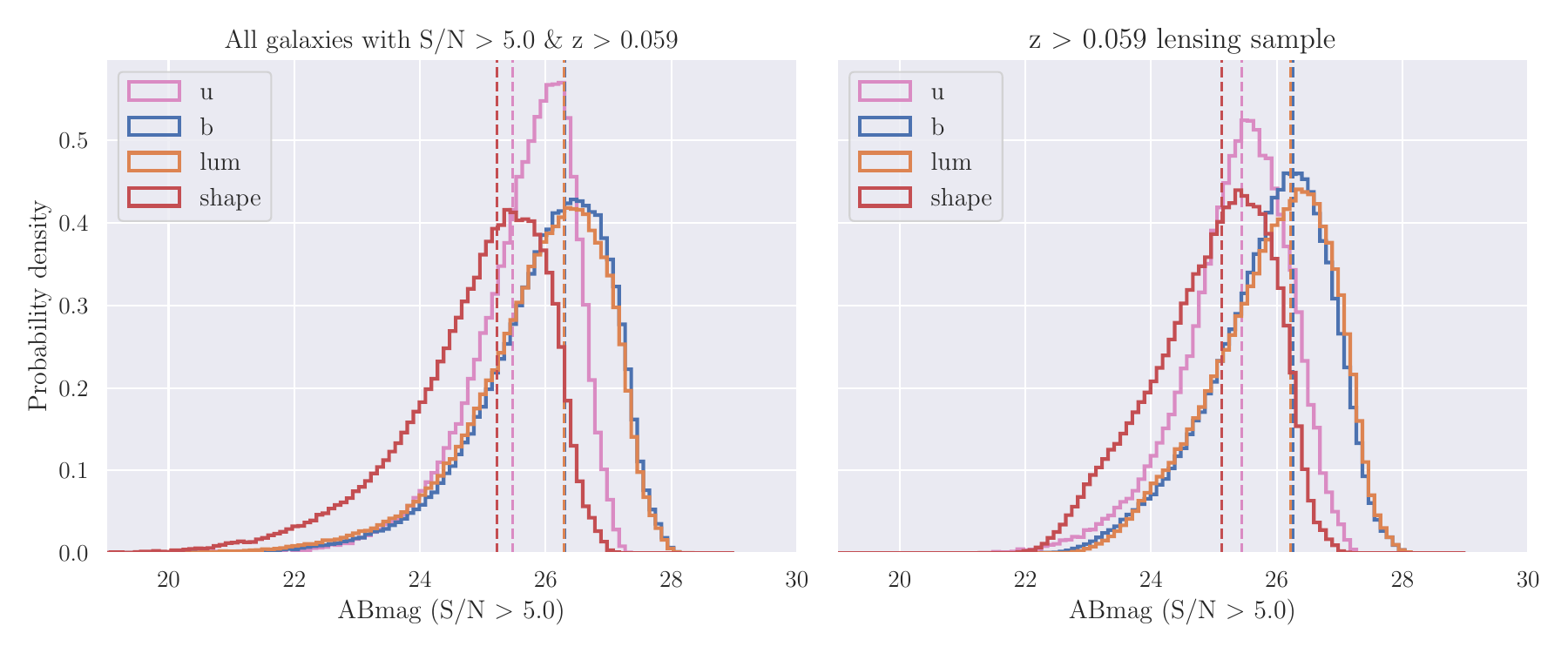}\\ 
   \includegraphics[width=0.9\textwidth]{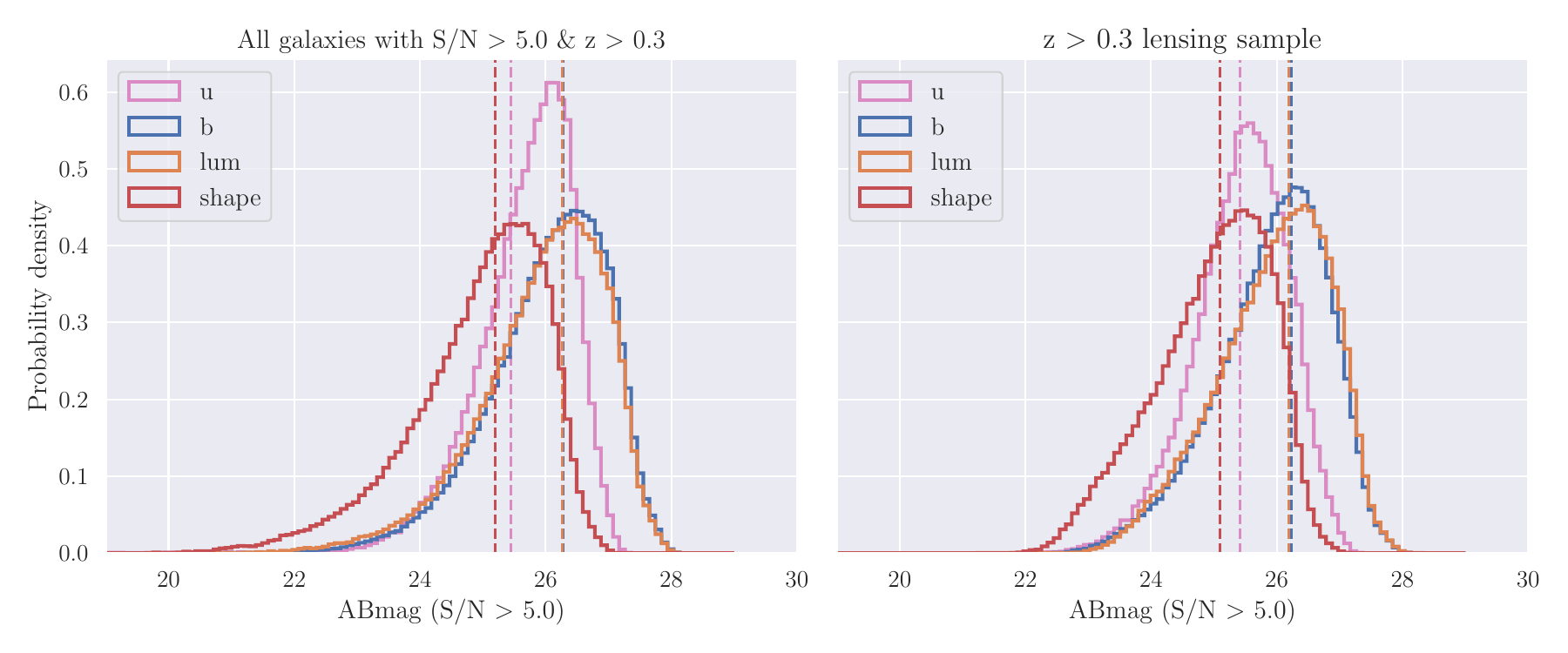}\\ 
   \includegraphics[width=0.9\textwidth]{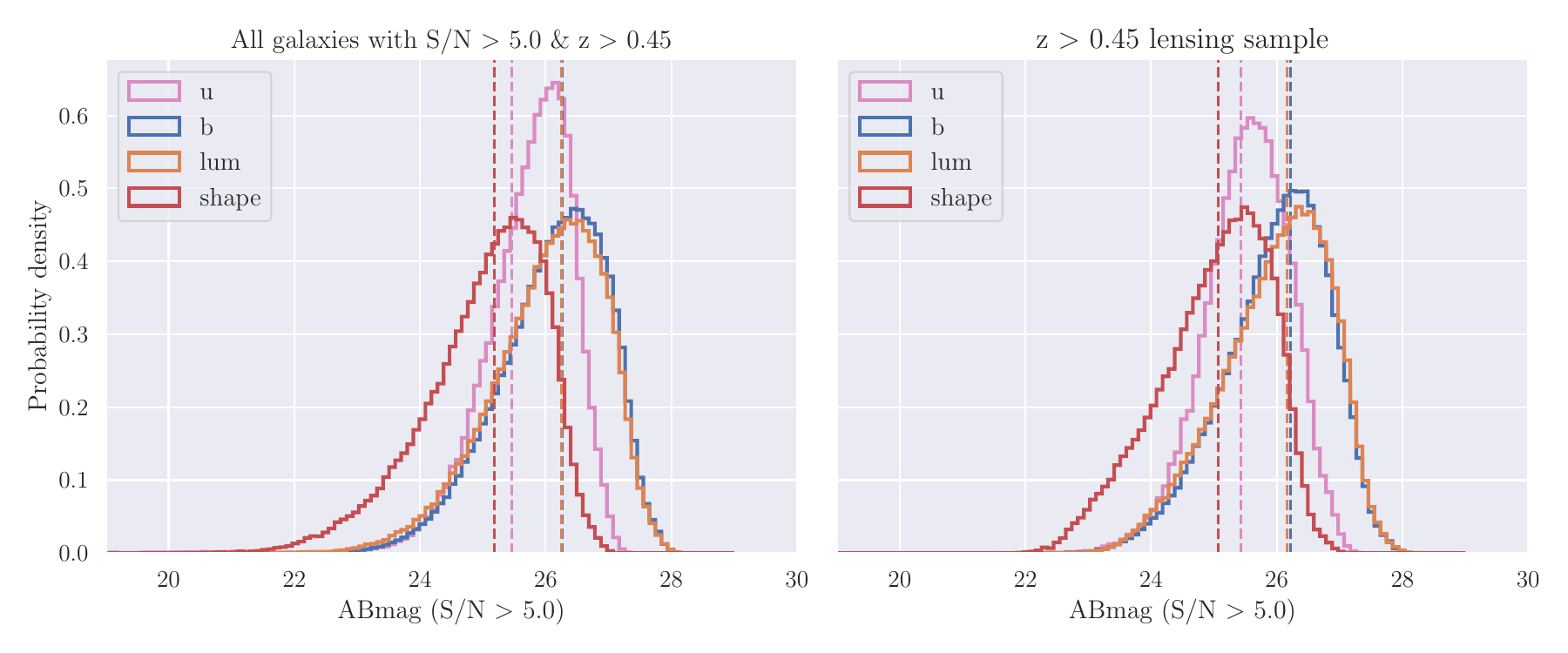}\\
   \caption{Galaxy brightness distributions shown as normalized histograms of galaxy counts, binned by AB magnitude. Dotted lines mark the $S/N=10$ depth. Left panels: all galaxies with S/N $>$ 5 and $z > z_{clust}$. Right: galaxies that pass lensing analysis selections including $z > z_{clust}$.}
   \label{fig:lum_funcs}
\end{figure}

Three hours of observation in \bmag yields a $\sim S/N=10$ depth of 26.3 before any lensing selections are made; lensing selections do not significantly change the depth. $S/N=10$ depths in \lum are similar to \bmag, while \umag and \shape depths are about a magnitude shallower. Values for all filters are listed in Table \ref{tab:summary}. 

Figure \ref{fig:lum_funcs} and the top panel of Figure \ref{fig:allgals_lumfunc_zdist} show galaxy brightness distributions, displayed as histograms of detected galaxy counts. Magnitudes are obtained from Source Extractor \texttt{FLUX\_AUTO} values, using the IMX455 detector gain and quantum efficiency to convert to the AB system. Distributions are shown for each of \umag, \bmag, \lum, and \shape, and the histograms are normalized such that the product of bin width and probability is one. 

The top panel of Figure \ref{fig:allgals_lumfunc_zdist} shows the distribution of the ``all galaxies'' sample with only a \SN{5} selection. Figure \ref{fig:lum_funcs} presents number counts as a function of brightness for galaxies behind clusters at $z=0.059$ (top row), $z=0.3$ (middle row), and $z=0.45$ (bottom row). The left panels show the $z_{gal} > z_{clust}$ sample for each cluster, and the right panels show the lensing-analysis galaxy samples. 

\subsection{Redshift distributions}\label{sec:results:redshifts}

\begin{figure}
\begin{center}
\includegraphics[width=0.8\textwidth]{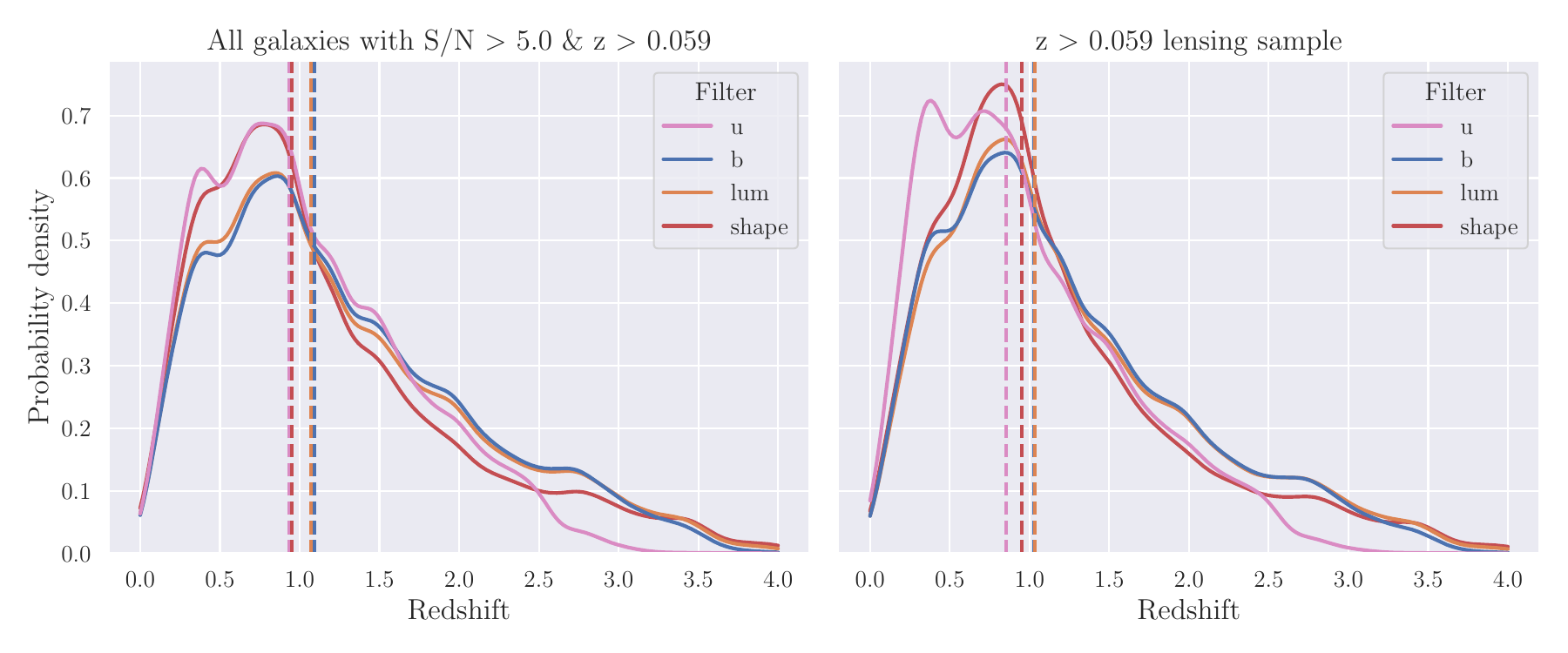}\\
\includegraphics[width=0.8\textwidth]{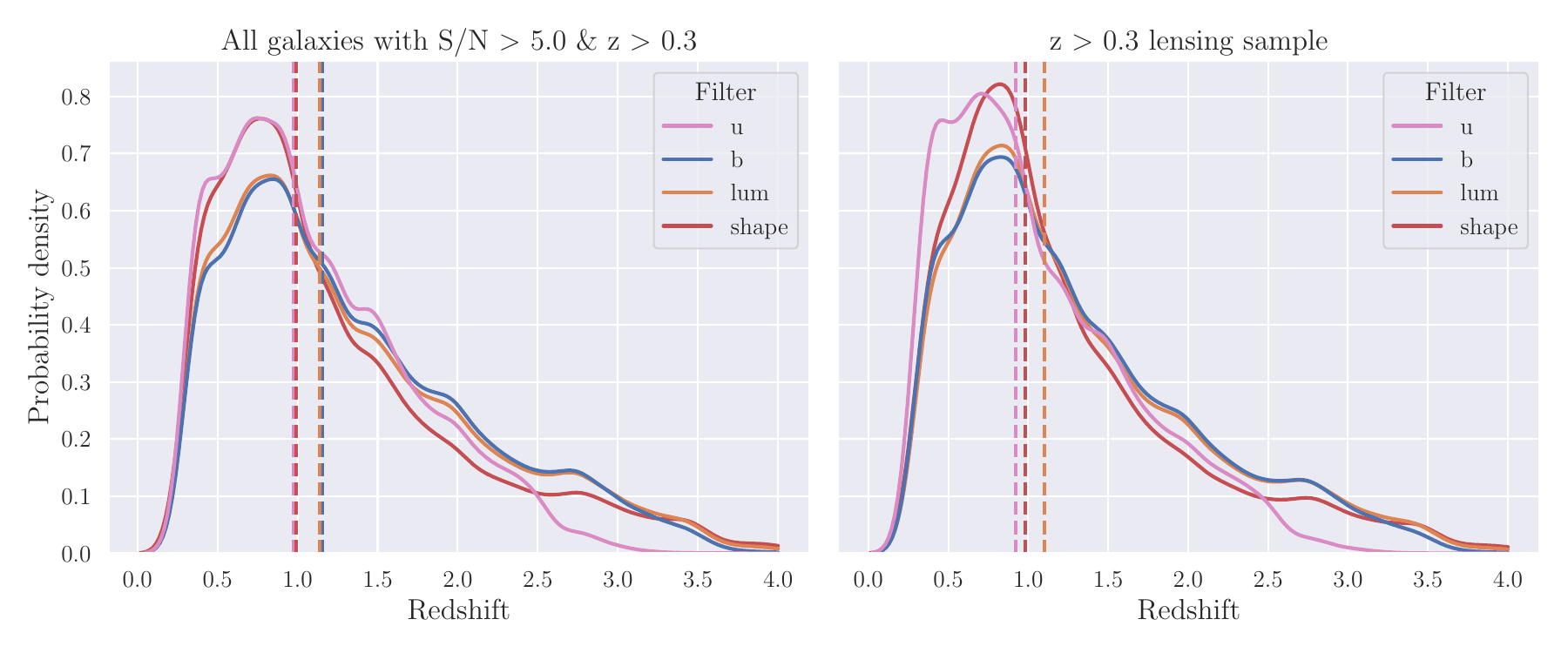}\\
\includegraphics[width=0.8\textwidth]{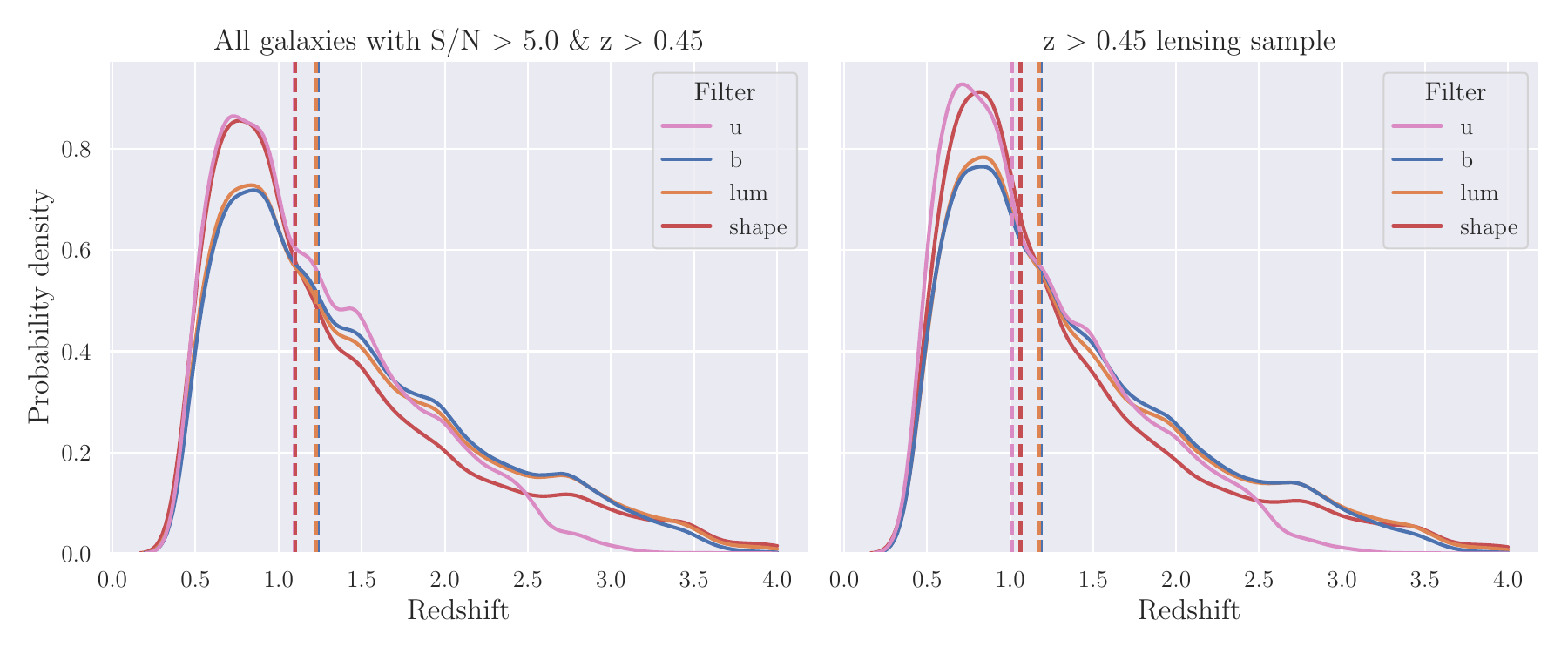}
\caption{Normalized galaxy redshift distributions behind a cluster at $z=0.059$ (top), $z=0.3$ (middle), and $z=0.45$. Left panels: all galaxies with S/N $>$ 5 and $z_{gal} > z_{clust}$. Right: galaxies that pass lensing analysis selections, including $z_{gal} > z_{clust}$.}
\label{fig:redshift_distributions}
\end{center}
\end{figure}

The strength of weak lensing signal depends on the relative distances of the cluster and background galaxies. Accordingly, we calculate galaxy redshift distributions (Figures \ref{fig:allgals_lumfunc_zdist}, bottom panel, and \ref{fig:redshift_distributions}). Distributions are obtained with kernel density estimation and are normalized to show relative probability density within a bandpass (the product of bin width and probability density equals unity). Dotted lines mark the median redshift in a given filter.

The bottom panel of Figure \ref{fig:allgals_lumfunc_zdist} shows redshift distributions for all galaxies detected in \umag (pink), \bmag (blue), \lum (orange), and \shape (red) coadds. The probability density in all bands peaks around $z=0.83$, with long tails past redshift $z=1.5$. 
Consistent with the depths in Section \ref{sec:results:depths}, \umag and \shape observations have mean and median redshifts about 0.15 units lower than the deeper bandpasses. 

As in Section \ref{sec:results:depths}, redshift distributions are shown both for a $z_{gal} > z_{clust}$ selection and for galaxies that pass all lensing analysis selections. The mean redshifts of background galaxies increases slightly with increasing cluster redshift, from a mean $b$ redshift of $\bar{z} = 1.3$ for $z_{clust} = 0.059$ to  $\bar{z} = 1.4$ for $z_{clust} = 0.45$. However, the changes are small, and lensing selections do not appear to change the mean or median background galaxy redshifts. The mean and median values of redshift in all bandpasses and galaxy samples are summarized in Table \ref{tab:summary}.  

\begin{figure}
\begin{center}
\includegraphics[width=0.8\textwidth]{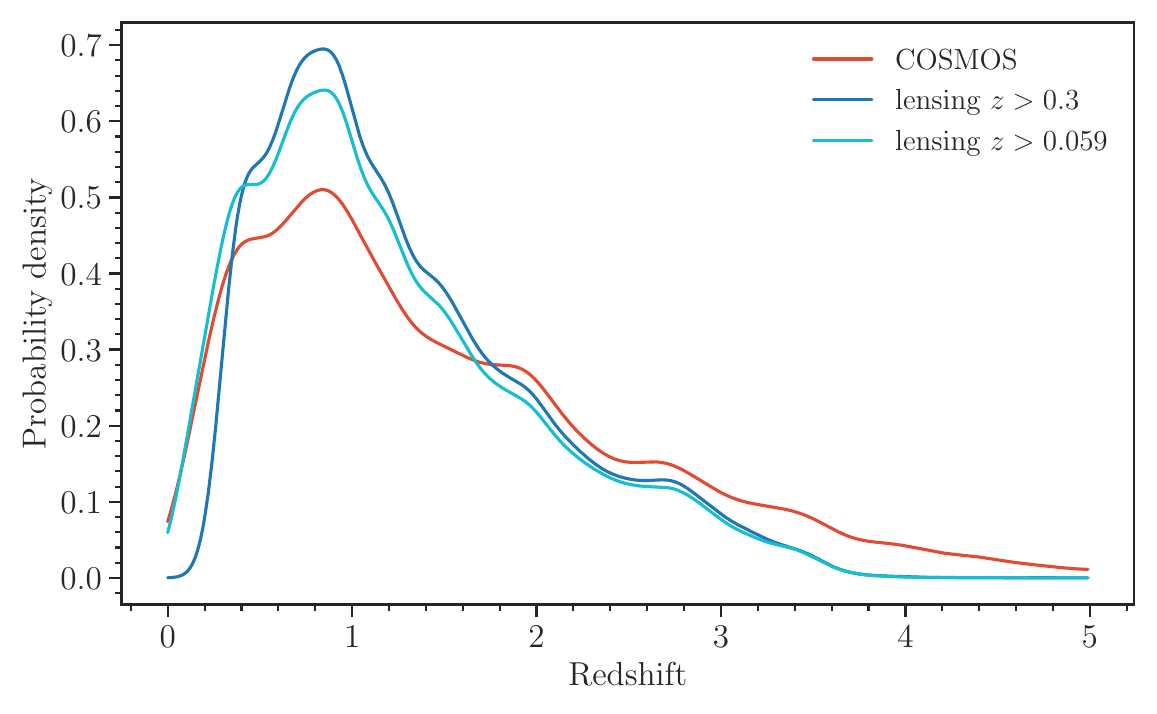}\\
\caption{Normalized redshift distributions of the COSMOS 2015 simulation input catalog (red line), the $z>0.3$ lensing sample (blue line), and the $z>0.059$ lensing sample (cyan line). The two lensing samples have lower mean redshifts ($\bar{z} = 1.2$ and $\bar{z} = 1.3$) than the COSMOS 2015 catalog ($\bar{z} = 1.5$).}
\label{fig:cosmos_redshifts}
\end{center}
\end{figure}

It is reasonable to ask whether the redshift distributions of the galaxy lensing samples are actually distinct from the input COSMOS 2015 catalog. We compare these in Figure \ref{fig:cosmos_redshifts}, which compares the redshift distributions of the COSMOS 2015 simulation input catalog and the lensing-analysis catalogs of \bmag observations of clusters with $(M,\,z) = (4.1 \ee{14}\, M_{\odot} \, h^{-1}, \, 0.059)$ and $(4.1 \ee{14}\, M_{\odot} \, h^{-1}, \, 0.3)$. The probability densities of both the COSMOS 2015 catalog and \superbit lensing samples are maximized at $z \sim 0.9$. However, the COSMOS 2015 catalog has a higher probability density at $z>1$. The mean redshift of the COSMOS 2015 catalog is $\bar{z}=1.5$, compared with $\bar{z} = 1.2$ for the $z>0.059$ lensing-analysis sample and $\bar{z} = 1.3$ for the $z>0.3$ lensing-analysis sample.

\begin{figure}
    \centering
    \includegraphics[width=0.63\textwidth]{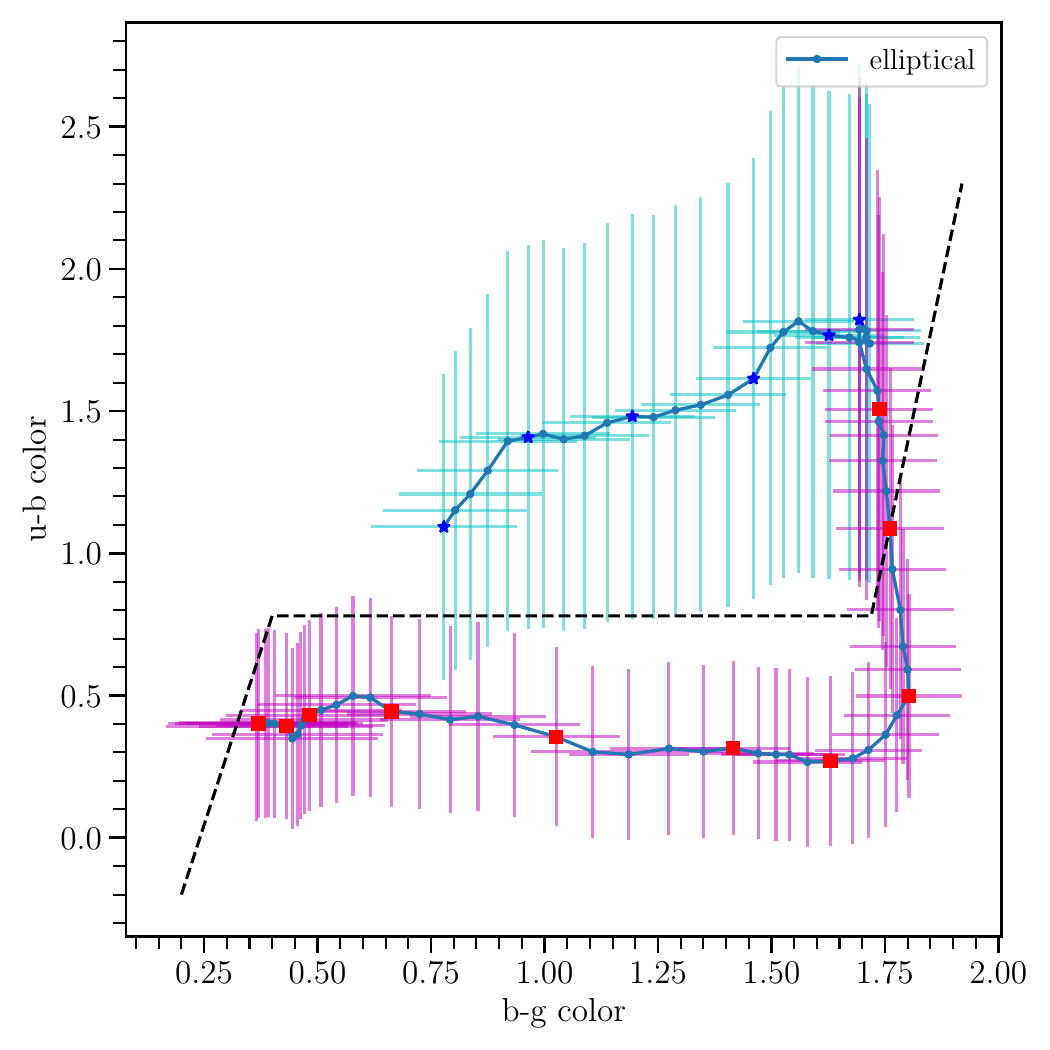}\\
    \includegraphics[width=0.63\textwidth]{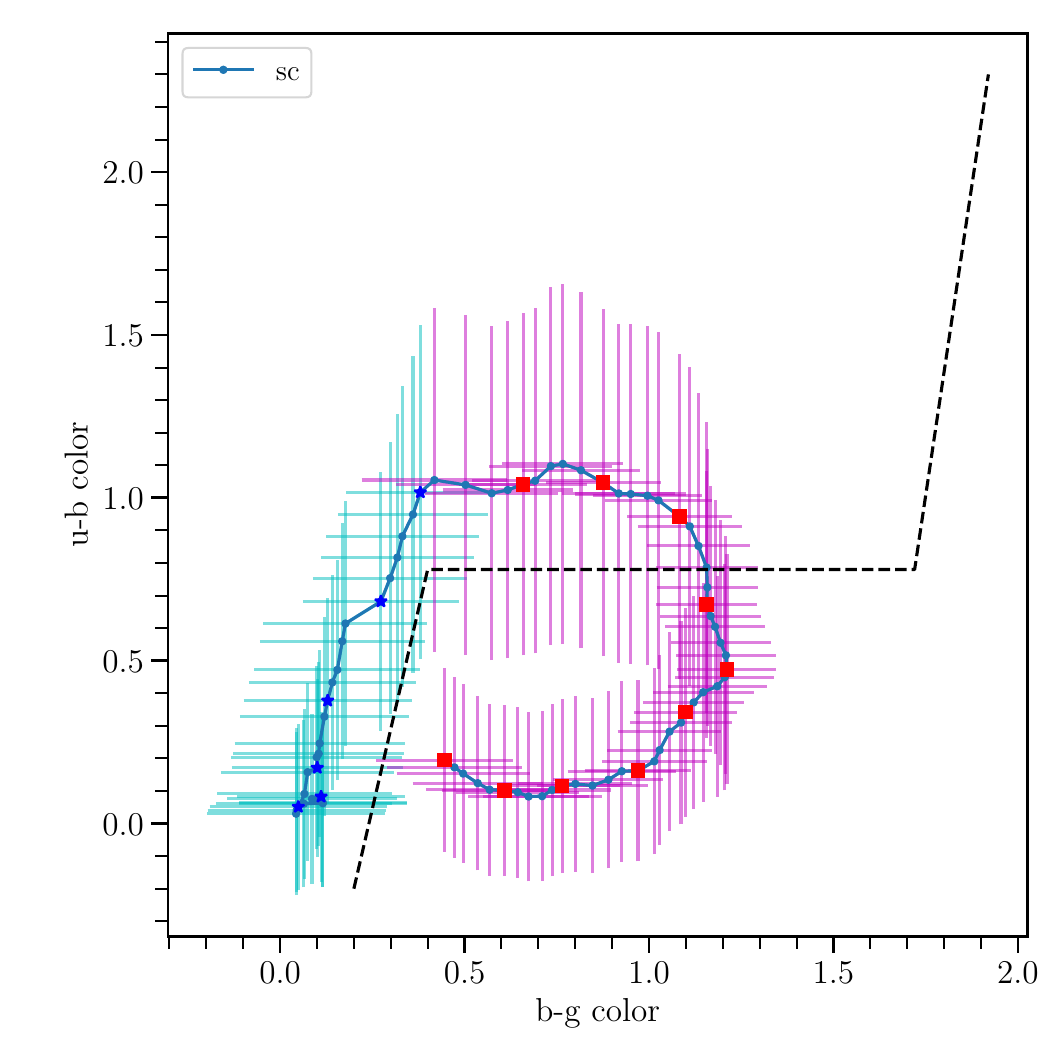}\\
    \caption{Redshift evolution of elliptical and spiral galaxies in $(b - g)$, $(u - b)$ color-color space. Blue solid lines show the observed galaxy color evolution between $0 < z_{\rm gal} < 1.5$; small points mark $\delta z = 0.02$ intervals. Large markers are included as a visual aid: blue stars and cyan error bars mark $z_{\rm gal} < 0.5$, red squares and magenta error bars mark $z_{\rm gal} \geq 0.6$. Error bars are 1 $\sigma$ color uncertainties for galaxies with $b$-band $S/N = 10$. Color-color tracks assume exposure times of 3 hr, 1.5 hr, and 3 hr in $b$-, $g$-, and $u$-bands, respectively. A `clean' color-color space, with low foreground galaxy contamination (no blue stars), suitable for a cluster at $z=0.5$, lies below the dashed black line.}
    \label{fig:colorcolor}
\end{figure}

The calculations above assume perfect knowledge of the redshift. In real \superbit observations, we will separate background (lensed) galaxies from foreground (unlensed) galaxies with galaxy color cuts. To optimize the exposure time per bandpass for an effective foreground/background separation, we investigated the evolution of ($u - b$), ($b - g$) and ($g-r$) colors with redshift for a range of galaxy types.

We sampled galaxy redshifts in the range $0 < z_{\rm gal} < 1.5$ at $\delta z = 0.02$ intervals for spectral templates from elliptical to starburst \citep{1996ApJ...467...38K}.
For each $\delta z = 0.02$ point, we transformed the galaxy's spectral energy distribution (SED) to the desired redshift and scaled the SED flux to achieve an integrated $S/N$ of 10 in the $b$-band filter, representing the minimum $S/N$ for inclusion in lensing analysis. Based on the scaled SED flux, we calculated the $b$-, $g$-, and $u$-band magnitudes along with their respective magnitude errors. By calculating uncertainties for a galaxy with $S/N = 10$, we obtained conservative error bars that allowed us to define realistic color-cut boundaries for our galaxy selection.

For a fiducial cluster redshift of $z = 0.5$, we determined that 3 hours of integration time in $b$, 1.5 hours in $g$, and 3 hours in $u$ provided optimal separation for galaxies of most spectral types. Figure \ref{fig:colorcolor} illustrates the color evolution for two spectral types (elliptical and disk-dominated spiral), depicted by solid lines. The small points represent the galaxy colors calculated at each $\delta z = 0.02$ interval. To aid the reader, we highlight specific foreground ($z_{\rm gal} < 0.5$) and background ($z_{\rm gal} \geq 0.6$) locations as blue stars and red squares, respectively, at intervals of $\delta z = 0.1$.

Error bars in Figure \ref{fig:colorcolor} represent predicted 1 $\sigma$ color uncertainties for the aforementioned exposure times and a galaxy $b$-band $S/N = 10$. Cyan error bars correspond to galaxies in the foreground of a $z = 0.5$ cluster, while magenta error bars indicate galaxies behind the cluster ($z_{\rm gal} > 0.6$). The dashed black lines in Figure \ref{fig:colorcolor} demarcate a `clean' color-color space for a $z=0.5$ cluster. The galaxy sample below the black lines is dominated by background galaxies at $z_{\rm gal} > 0.6$, with minimal contamination from foreground galaxies ($z_{\rm gal} \leq 0.5$).

\subsection{Mean shear profiles}\label{sec:results:shearprofiles}

As part of the pipeline validation effort, we also produce weak gravitational lensing shear profiles for all cluster realizations. Two examples of single-realization cluster shear profiles were shown in Figure \ref{fig:MiscShearProfiles1}. 

\begin{figure}
    \centering
    \includegraphics[width=0.95\textwidth]{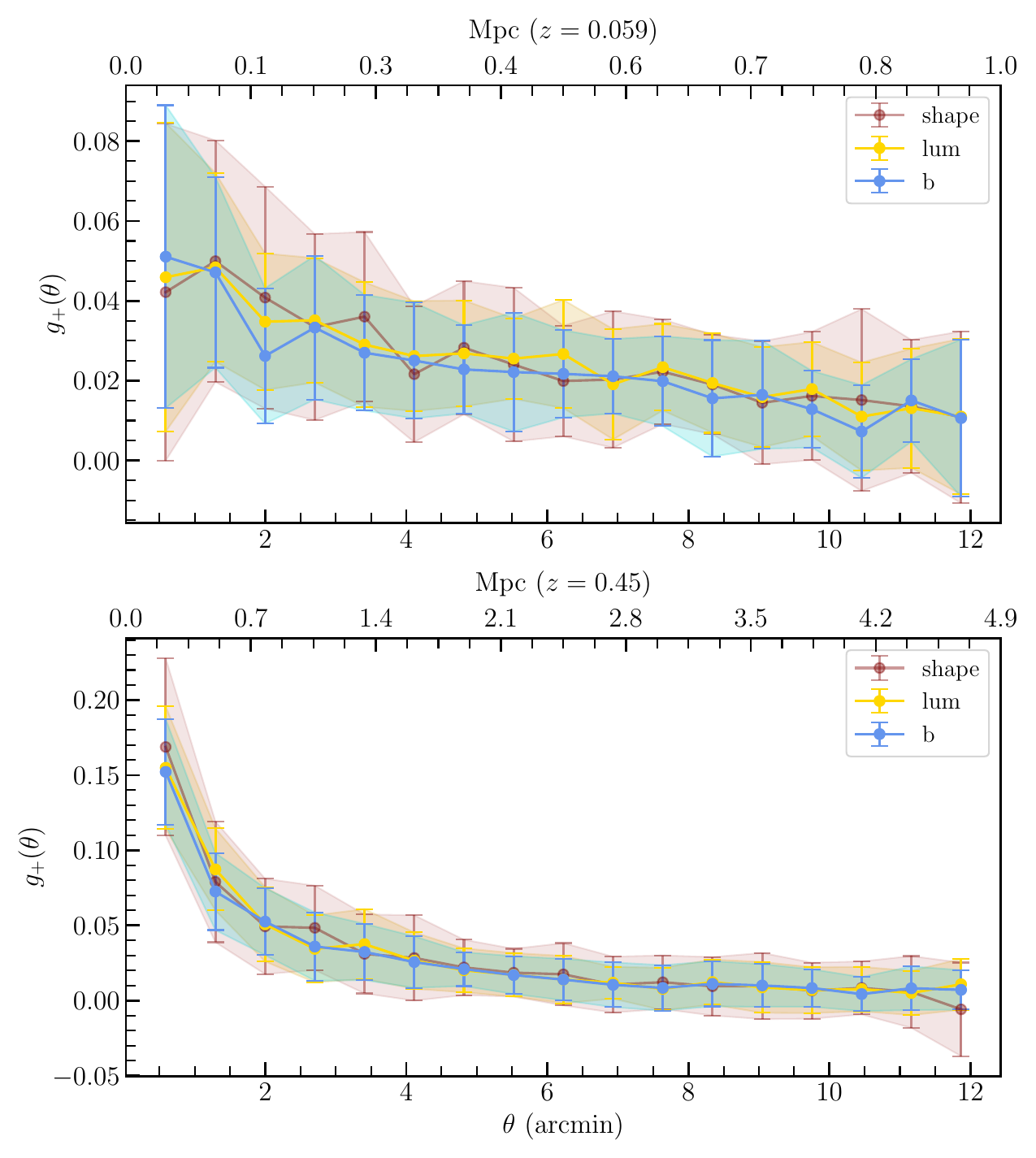}    
    \caption{Average tangential shear profile for 30 simulated galaxy clusters with identical masses and redshifts. Each point represents the mean value of the tangential shear \textit{profiles} across the cluster sample; error bars represent the standard deviation of the mean in each radial bin and illustrate the fluctuation of individual cluster measurements. Profiles are shown for observations in \bmag, \lum, and \shape (3 hours of exposure time each). Top: clusters with $M=4.1 \times 10^{14}\ M_{\odot} \, h^{-1}, \, z=0.059$. Bottom: clusters with $M=4.1 \times 10^{14}\ M_{\odot} \, h^{-1}, \, z=0.45$. The bottom X-axes show distance from the cluster center in arcminutes, while top axes are scaled to show the projected distance from the cluster center in kiloparsecs. The tangential shear profiles are easily detected in \superbit \bmag and \lum observations, shown in blue and gold, with no obvious advantage for \shape-band measurements (shown in red).}
    \label{fig:threecolor}
\end{figure}

To examine the claim that \superbit is capable of weak lensing measurements in blue bandpasses, we compare the mean tangential shear profiles of cluster observations in \bmag (\superbit's intended filter for galaxy shape measurement), \lum, and the Euclid VIS-like \shape filter in Figure \ref{fig:threecolor}. The mean tangential shear profiles of 30 realizations of $z=0.059$ clusters are shown in the top panel and $z=0.45$ clusters in the bottom panel. Each point represents the mean value of the cluster tangential shear \textit{profiles}, while error bars show the standard deviation of the mean in each radial bin. 

We find that the tangential shear profiles are easily detected in all three \superbit bandpasses. No differences in the mean values for \bmag, \lum and \shape are readily apparent for either cluster in Figure \ref{fig:threecolor}. Qualitatively, the \shape band error bars appear slightly larger than the \lum and \bmag error bars, which is consistent with the lower \shape source densities in Table \ref{tab:summary}. 

We emphasize that the shear profiles of Figure \ref{fig:threecolor} are averages of averages, and would not be used for shear calibration or mass fitting. Instead, the figure highlights the variability and reliability of the measured tangential shear  across the sample of clusters.

\section{Discussion}\label{sec:discussion}
    

We provide some additional commentary on our analyses and results here.

\subsection{Simulation inputs and effect on forecasts}\label{sec:discussion:simulations}

The simulations presented in Section \ref{sec:the_simulations} and which form the basis for Section \ref{sec:results} have many realistic features: an NFW cluster weak lensing profile, star flux and densities from Gaia coverage of \superbit targets, measured stratospheric sky brightnesses from \cite{gill2020optical}, and real galaxy redshifts and luminosities from COSMOS catalogs transformed to \superbit bandpasses. 

Although the simulated observations incorporate considerable complexity, there are a few limitations. First, they use a Gaussian approximation of the \superbit PSF. In reality, the space-like \superbit PSF features Airy rings and diffraction spikes (cf.\ Figure \ref{fig:psf}). 

We also do not model uncertainties of galaxy redshifts. This was a deliberate choice, as the systematic errors that redshift uncertainties introduce to weak lensing analysis are orthogonal to the pipeline validation aspect of this work and the shear calibration in forthcoming efforts. Even if we did attempt to incorporate redshift uncertainties, \superbit's strategy for determining redshifts may evolve as the campaign progresses, rendering such forecasting estimates moot. 

In addition, the validity of our forecast is limited by the simulation input catalog. There are few deep, high resolution observations in the blue and near-UV. A workaround is presented in Section \ref{sec:the_simulations}, but assumes that galaxy morphology parameters in the ($\bar{z} = 0.9$) GalSim-COSMOS catalog can be extrapolated to the ($\bar{z} = 1.5$) galaxies in the COSMOS 2015 photometric catalog. A more theory-driven approach could involve hydrodynamical simulations. However, the morphology of intermediate- to high-$z$ galaxies is itself a very active area of research. On balance, the high accuracy of the galaxy fluxes and realistic redshift distributions in our input COSMOS 2015 catalog outweigh any uncertainty in galaxy shapes. A full treatment of the galaxy catalog will be presented in a forthcoming paper by A.\ Gill et al.\ (in preparation).  

Finally, the input Gaia star catalogs are incomplete, as illustrated by the gap in the stellar locus of Figure~\ref{fig:piffstar_sizemag}. We do not believe that the dearth of faint stars affects our conclusions, as very faint stars near the ``zone of confusion'' would be excluded from PSF fits anyway. Future simulations will incorporate theoretical \texttt{TRILEGAL} star distributions.  

\subsection{Estimated observation depths, source densities, and redshifts}
 
A major goal of this analysis was to quantify the effect of weak lensing selections on galaxy number density. Table \ref{tab:summary} shows that weak lensing analysis selections cause a more significant decrease in source density (30-40\%) than redshift selections alone. The addition of lensing selections does not appear to significantly change the mean and median redshifts of the samples any more than a redshift cut alone.

A surprising result of Section \ref{sec:results:depths} is the high depth and source density in $u$. The deep NUV CLAUDS survey \citep{sawicki2019cfht} provides one of the few points of comparison for our own \umag findings. At a similar depth to ours (25.5 mag), they report a $5\ \sigma$ source density of $\log_{10} N = 4.58/{\rm deg}^{2}/$0.5 mag, or 10.7 galaxies per arcmin$^{2}$. This is 50\% lower than our maximum reported value of 15.4 galaxies per arcmin$^{2}$. The change in source density with redshift is also noteworthy: we report a 27\% decline in $u$ source density from $z=0.059$ to $z=0.45$, while over approximately the same redshift range, the CLAUDS survey reports a decline of $\sim 12\%$ \citep{2020MNRAS.494.1894M}. 

One possibility for the divergence is \superbit's smaller PSF: the CLAUDS survey experienced an average PSF FWHM of $0.92\arcsec$, but the \superbit \umag PSF FWHM is about $0\farcs278$. A smaller PSF translates to a higher source density, as objects that might otherwise be blended or smeared out over noisy pixels become resolvable. A more likely explanation is that our UV luminosities do not account for foreground extinction by Milky Way dust, which is significant in the UV and will certainly depress \umag source counts in real \superbit observations. If the GalSim-COSMOS shape parameters cannot be extrapolated to bluer bands and fainter galaxies, it is also possible that the galaxy morphologies in our catalog are inaccurate for \umag observations. The ultimate calibration for our simulations will be provided by the analysis of real \superbit observations in \umag and \bmag. 

\subsection{Impact on observation strategy}

Figures \ref{fig:joined_match}, \ref{fig:ann_match}, and \ref{fig:threecolor} show that the source density achieved in three hours of observation in \bmag or \lum is completely adequate for shear profile measurements. Observations in $b$ or $lum$ longer than three hours would confer limited advantages at a high cost in integration time (see Equation \ref{eqn:sourcedensity}). In fact, future analysis may reveal that shorter integration times would suffice, saving time during flight and allowing a greater number of targets to be observed. 

The final observation strategy will depend on the results of ongoing optics-on (jitter+optics) simulations in all \superbit bandpasses as well as a redshift analysis that is currently underway. However, Figures \ref{fig:joined_match} and \ref{fig:ann_match} strongly support the conclusions of \cite{2022AJ....164..245S} that \lum and \bmag observations are both faster and deeper than the Euclid VIS-like \shape when observing from the stratosphere. Our estimated source densities in these bandpasses also agree with \cite{2022AJ....164..245S} within uncertainties. 

Finally, Figure \ref{fig:threecolor} shows the feasibility of measuring galaxy cluster weak lensing signal in \bmag and \lum and that a broadband red filter like \shape offers no noticeable advantage over the bluer filters. This result supports our planned observing strategy of deep $b$ observations for galaxy shape measurements. 


\section{Conclusions and outlook for 2023}\label{sec:conclusions}
    

In this work, we have presented a first iteration of the galaxy shape measurement pipeline for \superbit's weak lensing analysis. The software and algorithms we employ---GalSim, SExtractor, PIFF, Metacalibration, NGMix---have been rigorously tested and were intended for widespread adoption by the community. Processing simulated observations has allowed us to test their implementation in this pipeline. Several years after the release of these tools, there is now a growing number of pipelines similar to ours, e.g. \texttt{ShapePipe} \citep{2022arXiv220404798G} and \texttt{run\_steps} \citep{2022ApJ...933...84F}, with more likely to come. 

Beyond pipeline validation, our simulated observations and catalogs provide estimates for the expected number density, depth, and redshift distribution of galaxies in deep, stratospheric imaging. We predict that \superbit can attain a depth of 26.3 mag in the \bmag filter and 25.5 mag in the \umag filter -- competitive with even the deepest ground-based surveys. We also find a total source density greater than 40 \Neff in three hours of integration time in both the \bmag and \lum bands. The source density remains high even after the application of lensing catalog selections: 25--30 \Neff in the \bmag bandpass. We expect that instrumental effects (including the optical PSF) will depress the source density. However, the \textit{relative} performance of \bmag, \lum, and \shape is unlikely to be affected and supports \superbit's observation strategy. 

This work also offers a look at the weak lensing tangential shear profiles expected for \superbit cluster observations, further confirming \superbit's capacity for weak gravitational lensing measurements in the blue. As with the other forecast survey properties, these weak lensing profiles are based on Gaussian approximations to the \superbit PSF and do not include redshift uncertainties. The vagaries of real observations will add some scatter to the final weak lensing measurements. Even with these caveats, the relative performance of different filters also supports \superbit's observation strategy.  

The \superbit pipeline and simulations remain in active development. Forthcoming improvements include source detection on a multi-bandpass composite image; galaxy shape measurement with the full \superbit PSF; inclusion of faint stars in simulated observations using stellar population synthesis models; and the addition of redshift uncertainty to the input galaxy catalog. While the pipeline includes tools for shear calibration, we do not validate them here. Instead, a complete shear calibration analysis will be presented in a forthcoming paper by S. Everett et al. (in preparation).

Though our pipeline has been developed specifically for \superbit weak lensing measurements, it is generic and can be refactored for weak lensing observations with other instruments. An obvious example is \superbit's successor mission, \textsc{GigaBIT}: a planned 1.3 m gigapixel class balloon-borne observatory \citep{10.1117/12.2630356}. Future pipeline developments will facilitate forecasting and survey planning for both \superbit and \textsc{GigaBIT}. 

Since the initial submission of this paper, we are excited to announce the successful launch and completion of the \superbit mission, which spent 40 days at float. The data calibration process is currently underway, and we will subsequently conduct an analysis along the lines described in this paper.

\superbit offers a new data product: wide-field, diffraction-limited $\lambda <$ 600 nm imaging deep enough to enable galaxy cluster weak lensing analysis. Our forecast galaxy number density and redshift distribution confirm \superbit's capability for weak lensing mass measurement in blue wavelengths. This demonstrates that even in the era of multi-billion-dollar space telescopes like JWST, \textit{Roman}, and \textit{Euclid}, nimble and low-cost missions like \superbit offer immense scientific potential and a complementary paradigm for space-based scientific observations. 

\vspace{5mm}

\software{Astropy \citep{astropy:2013, astropy:2018},
          GalSim \citep{ROWE2015121},
          Source Extractor \citep{1996A&AS..117..393B},
          NGMIX \citep{sheldon2015ngmix}, 
          Seaborn \citep{Waskom2021}, 
          Matplotlib \citep{Hunter:2007}
          }

\section{Acknowledgements}

Support for the development of \superbit is provided by NASA through APRA grant NNX16AF65G. 
Launch and operational support for the sequence of test flights from Palestine, Texas are provided by the Columbia Scientific Balloon Facility (CSBF) under contract from NASA's Balloon Program Office (BPO). Launch and operational support for test flights from Timmins, Ontario are provided by the \textit{Centre National d'\'Etudes Spatiales} (CNES) and the \textit{Canadian Space Agency} (CSA).

JR, EH, and SE are supported by JPL, which is run under a contract by Caltech for NASA. Canadian coauthors acknowledge support from the Canadian Institute for Advanced Research (CIFAR) as well as the Natural Science and Engineering Research Council (NSERC). The Dunlap Institute is funded through an endowment established by the David Dunlap family and the University of Toronto. UK coauthors acknowledge funding from the Durham University Astronomy Projects Award, STFC [grant ST/P000541/1], and the Royal Society [grants UF150687 and RGF/EA/180026].

The simulation input catalog is based on data products from observations made with ESO Telescopes at the La
Silla Paranal Observatory under ESO programme ID 179.A-2005 and on data products produced by TERAPIX and the Cambridge Astronomy Survey Unit on behalf
of the UltraVISTA consortium.

\bibliography{main.bib}{}
\bibliographystyle{aasjournal}

\end{document}